\documentclass[12pt,floatfix,onecolumn,notitlepage,aps,prb]{article}

\usepackage[fleqn]{amsmath}
\usepackage{amsfonts}
\usepackage{amssymb}
\usepackage{hyperref}
\usepackage{setspace}

\usepackage[title]{appendix}

\usepackage{epsfig}
\usepackage{geometry}
\usepackage{ctable}

\usepackage{float}

\newcommand{\cC}{{\mathcal{C}}}
\newcommand{\cO}{{\mathcal{O}}}

\newcommand{\rV}{{\boldsymbol{r}}}

\newcommand{\Cset}{{\mathbb{C}}}

\newcommand{\Rset}{{\mathbb{R}}}
\newcommand{\Sset}{{\mathbb{S}}}

\newcommand{\cV}{{\mathcal V}}
\newcommand{\cE}{{\mathcal E}}

\newcommand{\bomega}{{\boldsymbol{\omega}}}

\begin{document}

\title{On the global minimum of the classical potential energy for clusters bound by many-body forces}

\author{\sc{Michael K.-H. Kiessling}$^{1,*}$ and \sc{David J. Wales}$^2$\\
    $^1${\small{Department of Mathematics,
	Rutgers, The State University of New Jersey}}\\
     {\small{110 Frelinghuysen Rd., Piscataway, NJ 08854, USA}}\\
     {\small{email: miki@math.rutgers.edu}}\\
   $^2${\small{Yusuf Hamied Department of Chemistry, Lensfield Road, Cambridge CB2 1EW, UK}}\\
       {\small{email: dw34@cam.ac.uk}} }
\date{Revised version of November 26, 2023\vspace{0truecm}} 

\maketitle
\vspace{-1truecm}

\begin{abstract}\noindent
 This note establishes, first of all, the monotonic increase with $N$ of the average $K$-body energy of classical $N$-body 
ground state configurations with $N\geq K$ monomers that interact solely through a permutation-symmetric $K$-body potential,
for any fixed integer $K\geq 2$. 
 For the special case $K=2$ this result had previously been proved, and used successfully as a test criterion for 
optimality of computer-generated lists of putative ground states of $N$-body clusters for various types of pairwise interactions.
 Second, related monotonicity results are established for $N$-monomer ground state configurations whose monomers interact 
through additive mixtures of certain types of $k$-meric potentials, $k\in\{1,...,K\}$, with $K\geq 2$ fixed and $N\geq K$. 
 All the monotonicity results furnish simple necessary conditions for optimality that any pertinent list 
of computer-generated putative global minimum energies for $N$-monomer clusters has to satisfy.
 As an application, databases of $N$-body cluster energies computed with an additive mix of the dimeric Lennard-Jones 
and trimeric Axilrod--Teller interactions are inspected.
 We also address how many local minima satisfy the upper bound 
inferred from the monotonicity conditions, both from a theoretical and from an empirical perspective.
\end{abstract}

\vfill

\hrule
\smallskip
$^*$ {\footnotesize{Corresponding Author.}}

\smallskip
\copyright {\footnotesize{(2023) The authors. Reproduction, in its entirety, for non-commercial purposes is permitted.}}
\smallskip
\hrule
\smallskip

{\small{Accepted on November 29, 2023, for publication in \textbf{Journal of Statistical Physics}.}}

\section{Introduction}

  The quest for the lowest potential energy configurations of classical $N$-body clusters is pursued in various branches
of physics, chemistry, and biology; see \cite{Sugano91,Jellinek99,WalesBOOK,RoyBook} for an introduction to, and survey of, 
cluster research,
and see \cite{CCD} for databases of putative global minima discovered through computer experiments.
 The vast majority of these structures will possibly never be known for sure to be optimal, because
this problem is classified as NP-hard
\cite{WilleVennik}, \cite{Adip} in the field of computer science.
  Hence it is very likely, though not yet rigorously established, that there is no deterministic algorithm
for finding the global minimum with only polynomially (in $N$) many steps, for any $N$. 
 Thus, only for small $N$ can one be confident that the available computer algorithms will find the optimal configuration. 
In fact, the difficulty in locating the global minimum depends on the organisation of the underlying
energy landscape, which can change dramatically when the cluster size differs by a single 
monomer \cite{DoyeMillerWalesB,walesdmmw00,waless99}.
 A larger cluster that corresponds to a `magic number' size with a single funnel landscape may be
a relatively easy target, even though it possesses vastly more local minima than a smaller size cluster with
a double-funnel organisation \cite{DoyeMillerWalesA}.

 In this situation it is clearly useful to have necessary conditions for optimality that can be used to test
lists of putatively optimal data. 
 For a very large class of models in which the $N$ monomers in a cluster interact in a purely pairwise fashion, 
a necessary condition for optimality is the 
monotonic increase with $N$ of the average pair energy for the global minima.
 This result  was established in \cite{KieJCP}, after noting that earlier arguments of 
\cite{KieJSP} apply under considerably less restrictive assumptions than stipulated in \cite{KieJSP}.
 The test has revealed failures in computer-generated lists of putatively
optimal data; see \cite{KieJSP} for the generalized Thomson problem, and \cite{KieJCP} for the Lennard-Jones cluster problem.
 In \cite{KieJCP} data for clusters of water molecules were also tested, and no failures were detected.

 While this monotonicity test applies to the numerous
cluster models described by purely pairwise interactions, it cannot be used
to test putative global minima that involve irreducible $K$-body interactions with $K>2$, beginning with
$K=3$, where the Axilrod--Teller trimer interaction provides a physically relevant example \cite{AxilrodTeller}.
Such interactions are usually termed `many-body', as opposed to two-body.
 In this paper we explore whether similar necessary conditions for optimality exist for $N$-monomer
cluster models involving such many-body interactions.
 We begin with clusters whose monomers interact through $k$-body interactions with a single $k=K$, 
then generalize to mixtures with several $k\leq K$.

 In \S \ref{sec:K}  we prove the monotonic increase with $N$
of the optimal average $K$-body energy for clusters of $N\geq K$ monomers with irreducible $k$-body interactions 
for a single $k=K\geq 2$; when $K=2$, the proof reduces to the one in \cite{KieJCP}.
The number of local minima that may satisfy the
 implied test is also addressed in \S 2, providing some indication of how restrictive the bound can be in practice.
 In \S \ref{sec:MULTIk}  we generalize our discussion to cluster models for $N$ monomers that
interact with a mix of different $k$-meric interactions, $k\leq K$, for $K$ fixed.
 In \S \ref{sec:LJAT} we report the outcome of our test run on a family of putative global minima for $N$-body clusters that
interact with two-body Lennard-Jones plus three-body Axilrod--Teller potentials.
 We conclude in \S \ref{sec:FIN}.

\section{$N$-body clusters with a single $K$-body potential}
\label{sec:K}
\vspace{-.15truecm}

Our notation, and model assumptions, largely follows \cite{KieJCP}.
Thus, by $S\in {\mathfrak{S}}$ we denote a \emph{state} of a monomer, a point in a \emph{monomer state space} $\mathfrak{S}$.
 The position vector of component coordinates for a monomer is always part of the state, but there may be other variables,
e.g.~a unit vector for a polarized monomer's orientation in space, or some more complicated set of variables.

 By ${\cal C}^{(N)} := (S_1, ..., S_N)\in \mathfrak{S}^N$ we denote the \emph{ordered configuration} of an $N$-monomer cluster.
 The ordering is for notational convenience only and not of any intrinsic significance.
 It is implicitly understood that, if $N\geq 2$, then no two of the $N$ state variables in ${\cal C}^{(N)}$ coincide.

 For $K\geq 2$ fixed, by $V$ we denote a real-valued,  permutation-symmetric, 
irreducible $K$-state potential energy, also called ``$K$-body potential.''
 By ``permutation-symmetric'' we mean that 
$V(S_{i_1},...,S_{i_K})$ is invariant under all permutations of the $K$ state variables in the argument of $V$;
this definition takes care of the remark that the labeling of state variables in the configuration has no intrinsic significance.
 By ``irreducible'' we mean that $V$ cannot be written as a sum of $k$-body
potentials with $k< K$.
 
 For $N\geq K$ the potential energy of what we call an ``$N$-body cluster with monomers bound by a single $K$-body interaction'' is 
given by
\begin{equation}\label{WdefK}
W({\cal C}^{(N)} ) : = 
\sum\; \cdots\!\!\!\sum_{\hskip-1.2truecm 1\leq i_1<\cdots<i_K \leq N} V(S_{i_1},...,S_{i_K}).
\end{equation}
 It is assumed that for any $N\geq K$ there exists a globally minimizing $N$-monomer configuration 
${\cal C}^{(N)}_{\rm min}$ of $W$.
 Note that $K\geq 2$ is necessary for cluster formation, though not sufficient.
 Further assumptions are needed to guarantee that minimizers ${\cal C}^{(N)}_{\rm min}$ of $W$ are
genuine $N$-body clusters, but we do not need those assumptions to prove our monotonicity results 
for the optimizers ${\cal C}^{(N)}_{\rm min}$ of $W$.

 Next we define the average $K$-body energy of an $N$-monomer configuration by
\begin{equation}\label{eq:AveVK}
\langle V\rangle({\cal C}^{(N)}) := 
\genfrac{}{}{0.5pt}{0}{{}_1^{}}{\genfrac{(}{)}{0pt}{1}{N}{K}}
\sum\; \cdots\!\!\!\sum_{\hskip-1.2truecm 1\leq i_1<\cdots<i_K \leq N} V(S_{i_1},...,S_{i_K}),
\end{equation}
and the optimal average $K$-body energy as
\begin{equation}\label{optVKave}
v(N):= \min_{\cC^{(N)}} \langle V\rangle \big(\cC^{(N)}\big).
\end{equation}
 Since by assumption a minimizing configuration $\cC^{(N)}_{\rm min}$ exists, (\ref{optVKave}) can be written as
\begin{equation}\label{optVKaveREWRITE}
v(N) = \langle V\rangle \big(\cC^{(N)}_{\rm min}\big).
\end{equation}

This ends our list of definitions and assumptions.
We next list some examples for the important special case $K=2$ in \S \ref{sec:Thomson} to \S \ref{sec:water}, and then 
two examples for $K=3$ in \S \ref{sec:trimer} and \S \ref{sec:AT}. 

\subsection{Examples of models with a single $K$-body potential}

\subsubsection{The generalized Thomson problem}
\label{sec:Thomson}

For the generalized Thomson problem in $d+1$ dimensions  $\mathfrak{S} = \Sset^d\subset\Rset^{d+1}$, where
$S = \rV\in\Rset^{d+1}\cap\{|\rV| =1 \}$ is a unit-length position vector
in $d+1$-dimensional Euclidean space, representing the location of a generalized point charge on the unit sphere
in $\Rset^{d+1}$, with $d\geq 1$.
 The pair interaction of two charges can be written as (in reduced units of length and energy)
\begin{equation}\label{RieszV}
V(\rV_i,\rV_j) := \frac{1}{s}\left(\frac{1}{r_{ij}^s} -1\right),\quad s\in \Rset,
\end{equation}
with $r_{ij}^{}:=|\rV_i-\rV_j|\in (0,2]$ the $d+1$-dimensional Euclidean distance between the position vectors of the two charges.
 The parameter $s$ is known as the Riesz parameter. 
 Although the force for the pair potential $V(\rV_i,\rV_j)$ is repulsive for all $|\rV_i-\rV_j|>0$, the restriction that the
position vectors are unit vectors guarantees that a minimum of $W$ as defined in (\ref{WdefK}) exists on $\mathfrak{S}^N$ 
for all $N\geq 2$ and all $s\in\Rset$ --- provided one extends the interaction continuously to $r=0$ if $s < -2$.
 It should be noted that only when $s\geq -2$ do minimizers exist that are genuine $N$-point configurations.
 For $s< -2$ the minimizers with even $N$ are antipodal two-point configurations over which the $N$ position 
vectors are evenly distributed, while those with odd $N$ have their position vectors distributed over not more
than three points on a great circle \cite{Bjoerck}, but the solution when $N$ is odd is rigorously known only when $N=3$ \cite{KieYi}.
 
 Since the interaction (\ref{RieszV}) is invariant under the orthogonal group $\cO(d+1)$ that acts on $\Sset^d$, 
the $N$-body minimizers are obviously non-unique. 
 When listing (putative) optimizers $\cC^{(N)}_{\rm min}$ for the generalized Thomson problem it is tacitly 
understood that they represent a whole group orbit of $\cO(d+1)$.
 
 We highlight some special cases.
 The Thomson problem (so named originally in \cite{Whyte}) corresponds to the parameter choices $s=1$ and $d=2$,
i.e.~point charges that interact with repulsive Coulomb forces in $\Rset^3$ but are confined to $\Sset^2\subset\Rset^3$.
 Of separate interest is the case $s\to 0$ in (\ref{RieszV}), which gives $V(\rV_i,\rV_j) = \ln \frac{1}{r_{ij}^{}}$.
 This case, combined with $d=2$, features prominently in the original formulation of Smale's 7th problem \cite{Smale}.
 Another important example is $s=-1$.
 Both of these cases, $s\to 0$ and $s=-1$, are equivalent to the problem of maximizing a mean distance among $N$ points 
on the unit $d$-sphere in $\Rset^{d+1}$: the geometric mean distance when $s=0$ and the arithmetic mean distance when $s=-1$.
 When $d = 1$, the optimal solutions coincide.
 Interestingly, even for $d = 2$ the answers do not always agree --- they differ for the first time
at $N=7$; see  \cite{NBK}, formula (29) and the corresponding text.

 These examples are three noteworthy cases of interesting and related problems collected into a 
continuous single-parameter family known as the generalized Thomson problem.
  Introduced before the advent of quantum mechanics by J.~J.~Thomson \cite{Thomson04} to analyse 
the structures caused by electron-electron repulsion in atoms, the Thomson problem and its variations 
continue to furnish insights into questions of interest in condensed matter physics \cite{AltschulerETal}, \cite{BCNTa}, \cite{BCNTb}.

\subsubsection{Baxter's sticky hard sphere model}
\label{sec:Baxter}

Our second $K=2$ example is known as \emph{Baxter's sticky hard sphere model}.
 In this case $\mathfrak{S} = \Rset^d$, with $d\geq 1$, and $S = \rV\in\Rset^d$ is the position vector of the
center of a hard sphere  of radius $1/2$.
 The pair potential $V_2(S_i,S_j) := V_{\mbox{\tiny{shc}}}(|\rV_i-\rV_j|)$ is given by
\begin{equation}\label{Vshc}
V_{\mbox{\tiny{shs}}}(r) 
:= \begin{cases}\, \infty \quad \mbox{if} \quad  r < 1 
                  \\ \! -1 \quad \mbox{if} \quad r = 1
                   \\ \ 0 \quad\ \mbox{if} \quad r > 1
\end{cases}.
\end{equation}
 This potential corresponds to the following set-theoretical / point set topological narrative. 
 The monomers are  closed spheres of radius $1/2$ in Euclidean topology, with a pair energy 
that vanishes when the two spheres have empty intersection, while it is $-1$ if the intersection of the two spheres 
is a single point (a ``contact point''), and it is $\infty$ if their intersection is a closed 
set with non-empty interior.
 Thus, the globally energy-minimizing clusters of $N$ sticky spheres are identical with the 
configurations of $N$ spheres that feature the maximal number $c(N,d)$ of contact points.
Finding the global minimum for general $d$ and $N$ is a difficult open problem, 
yet only $d\leq 3$ would seem to be of direct relevance to chemical physics; see below.

 Of course, the $d=1$-dimensional version is quite trivial.
 The spheres can be identified with closed intervals (rods) of length 1,
and any $N$-body minimizer of the energy functional $W$ is trivially given by concatenating $N$ closed unit length 
intervals to form a single closed interval of length $N$.

 The $d=2$ dimensional version is not quite so trivial, yet all global minimizers have been determined \cite{Harborth}.
 The spheres can be identified with closed disks of radius $1/2$.
 Any global $N$-body minimizer $\cC^{(N)}_{\rm min}$ of the energy functional is 
a ``spiral array'' of such disks, and the optimal energy is given by
\begin{equation}
 W\big(\cC^{(N)}_{\rm min}\big) = - \lfloor 3N - \sqrt{12N - 3}\rfloor,
\end{equation}
where $\lfloor\ \cdot\ \rfloor$ is the floor function.
 When $N$ is large, every sphere in the interior of the configuration is in contact with six neighbors, giving the packing the 
hexagonal appearance of a honeycomb lattice.

 In $d\geq 3$ dimensions the solutions are more elusive.
As noted in \cite{BezdekKhan}, the global minimum $\cC^{(N)}_{\rm min}$ in $\Rset^3$ is known for $N\in\{2,3,4,5,13\}$.
 Yet, while for $N\in\{2,3,4,5\}$ (sometimes called the ``trivial cases'') the problem is straightforward to solve, 
it is worthy of note that for $N=13$ the solution is obtained from Newton's answer ``12'' to the famous ``kissing problem'' 
in $\Rset^3$ (\emph{What is the maximum number of congruent spheres in $\Rset^3$ that can simultaneously touch (viz.~``kiss'') 
a given sphere?}),
 where the rigorous proof required about 250 years of additional thought \cite{SchVdW}.
 
 A good overview of the state of the  art of rigorous results is \cite{BezdekKhan}.

In addition to the global minima, the growth in the number of distinct local minima
for the $d=3$ sticky hard sphere model has been investigated in several studies \cite{hoy_structure_2012,hoy15,holmes16}.
Furthermore, the correspondence between these local or global minima and those of Lennard-Jones-type clusters 
with different distance exponents has been analysed \cite{PhysRevE.97.043309}.
The structures can also be modelled using a Morse potential \cite{Morse29} with a very short range
parameter \cite{Wales10b}, which reflects the large pair equilibrium distance that results from the
excluded volume of the colloids.
 
In chemistry, the sticky hard sphere model is of interest in providing a simple potential for investigation of
complicated phenomena, such as phase changes, nucleation and growth \cite{TeoSloane,holmes13,holmes17,kallus17,hoy_minimal_2010}.
 Clusters of colloidal particles formed from polystyrene microspheres \cite{MengABM10,Crocker10} provide an
experimental realisation, and have been proposed as building blocks for the design of
structures with higher order organisation \cite{GlotzerS07,SacannaICP10,GlotzerE11,wolters2015self}.

\newpage

\subsubsection{Water Clusters}
\label{sec:water}

The third $K=2$ example we mention is the TIP5P \emph{water cluster model} \cite{MahoneyJ00}, for which 
$\mathfrak{S} = \Rset^3\times \Sset^2\times \Sset^1$,
where the first factor accounts for the position vector $\rV$ of the oxygen atom and the remaining 
two factors account for the three Euler angles $(\Theta,\Psi,\Phi)$ that fix the orientation of the rigid tetrahedral charge
distribution of the H$_2$O monomer w.r.t.~a convenient Cartesian reference frame.
 The TIP5P pair interaction is an additive mix of Lennard--Jones and Coulomb pair potentials, 
conveniently written (in reduced units of charge, length and energy) as
\begin{equation}
V(S,\widetilde{S}) :=
 \frac{1}{r^{12}} - \frac{1}{r^6}
+
\sum_{a \in\{1,...,4\}} \sum_{b \in\{1,...,4\}} \frac{q_a q_b}{r_{ab}},
\end{equation}
where 
$r:=|\rV-\tilde\rV|\in \Rset_+$ is the $3$-dimensional Euclidean distance between the position vectors of the two 
oxygen atoms, $r_{ab}\in (0,\infty)$
is the distance between the $a$-th charge on one H$_2$O monomer and the $b$-th charge on the other, with
$q_a$ and $q_b$ the respective charges on the corresponding monomers.
 The charges of a monomer satisfy $|q_a| = q$ and $\sum_{a}q_a=0$.
 We remark that once the states $S$ and $\widetilde{S}$ are given, the distances $r$ and $r_{ab}$ are determined.
 At small separations $r$ of the monomers the pair interaction is dominated by the $1/r^{12}$ term, while for large 
separations the pair interaction is asymptotic to $V_{\mbox{\tiny{dip}}}(\rV_i,\rV_j;\bomega_i,\bomega_j)$, where
\begin{equation}\label{Vdip}
V_{\mbox{\tiny{dip}}}(\rV_i,\rV_j;\bomega_i,\bomega_j) := 
\wp^2\frac{\bomega_i\cdot\bomega_j - 3 \cos\gamma_{i}\cos\gamma_j}{r_{ij}^3}
\end{equation}
is the familiar permanent dipole-dipole interaction potential, with $r_{ij}$ the  Euclidean distance 
between the two H$_2$O monomers, with  $\gamma_i\in[0,\pi]$ defined by 
$(\rV_i-\rV_j)\cdot\bomega_k = : r_{ij}\cos\gamma_k$ for $k\in\{i,j\}$, and with $\wp\bomega$ the permanent electric 
dipole moment of H$_2$O, where $\boldsymbol{\omega}\in\Sset^2$ represents a unit vector associated with the dipole.
 Clearly, $V_{\mbox{\tiny{dip}}}(\rV_i,\rV_j;\bomega_i,\bomega_j)$ is negative for suitable 
orientations of the two monomers, increasing to zero at infinite separation.
 This result guarantees the existence of a global minimum cluster configuration for (\ref{WdefK}) and 
each $N\geq 2$, with $W\big(\cC^{(N)}_{\rm{min}}\big)<0$.

If $q_a=0$, $a\in\{1,...,4\}$, then
the Euler angles are rendered irrelevant and the state space can be taken as 
$\mathfrak{S} = \Rset^3$, with $S = \rV\in\Rset^3$ the position vector of an atom.
The TIP5P interaction then reduces to the Lennard-Jones pair potential, 
\begin{equation}\label{eq:VLJ}
V_{\mbox{\tiny{LJ}}}(r) := \frac{1}{r^{12}} - \frac{1}{r^6},
\end{equation}
with $r=|\rV-\tilde{\rV}|\in (0,\infty)$ as defined earlier.

\subsubsection{An area analogue of the maximal average distance problem on $\Sset^2$}
\label{sec:trimer}

Our first $K=3$ example is a trimeric variation on the theme of the problem to maximize the arithmetic mean of all
dimer lengths; cf. \S \ref{sec:Thomson} with $s=-1$.
 Our state space is $\mathfrak{S} = \Sset^2\subset\Rset^{3}$, and
$S = \rV\in\Rset^{3}\cap\{|\rV| =1 \}$ is a unit-length position vector of a point particle on $\Sset^2\subset\Rset^{3}$.
 The trimeric interaction reads
\begin{equation}\label{eq:negAREA}
V(\rV_i,\rV_j,\rV_k)
 := - \tfrac12 \left| (\rV_k-\rV_j)\boldsymbol{\times}(\rV_i-\rV_j) \right|,
\end{equation}
i.e. $- V(\rV_i,\rV_j,\rV_k)$ is simply the area of the triangle in $\Rset^3$ with corners at $\rV_i$, $\rV_j$, $\rV_k$.
 The potential favors area-enhancement, but the compact configuration space $\mathfrak{S}^N$ guarantees non-trivial
minimizers of $W$ as given in (\ref{WdefK}) for each $N\geq 3$. 

We are unaware of any studies of this area analogue of the maximal-pairwise-distance problem on 
$\Sset^2$ \cite{Beck}. However, we note that maximizing the area of the smallest triangle among those spanned by the three-point 
subsets of $N\geq 3$ points in a convex two-dimensional set features in the celebrated Heilbronn problem 
\cite{Roth}, \cite{Goldberg}, \cite{KPS}.

\subsubsection{A trimerically stabilized Axilrod--Teller interaction}
\label{sec:AT}

Our second $K=3$ example is a stabilized Axilrod--Teller  interaction. 
In this case $\mathfrak{S} = \Rset^3$, and $S=\rV$ represents the position of the center of mass of a polarizable monomer.
 A monomer interacts irreducibly with two other monomers of the same kind through the trimeric potential
\begin{equation}
\label{eq:ATstab}
V(\rV_i,\rV_j,\rV_k) :=
 \frac{1}{r_{ij}^6r_{jk}^6r_{ki}^6}
+ 
 \frac{1+3 \cos\gamma_i \cos\gamma_j\cos\gamma_k}{r_{ij}^3r_{jk}^3r_{ki}^3}.
\end{equation}
 Here, $r_{jk} :=|\rV_j-\rV_k|$ is the  Euclidean distance between the position vectors
of monomers ${}_j$ and ${}_k$, and $\gamma_j\in[0,\pi]$ is defined by 
$(\rV_k-\rV_j)\cdot(\rV_i-\rV_j)= :r_{kj}r_{ji}\cos\gamma_j$.

The second term in (\ref{eq:ATstab}), usually referred to as the Axilrod--Teller potential,
or triple-dipole term, dominates at large distances and is obtained from third order quantum-mechanical perturbation 
theory  \cite{AxilrodTeller}.
 The first term is not usually considered, but is needed here to stabilize the interaction at short range
using only irreducible trimer potentials.
 The interaction (\ref{eq:ATstab}) may be viewed as a three-body analog of the pairwise Lennard-Jones potential (\ref{eq:VLJ}), 
where the $-1/r^6$ term that dominates at large distances can also be obtained from quantum-mechanical perturbation
theory, while the stabilizing $1/r^{12}$ term is a convenient form for the repulsive 
short-range part of the potential.

 Since the three angles sum to $\pi$, the interaction is negative for large $r_{ij}r_{jk}r_{ki}$
 whenever one of the angles is $\approx \pi$, positive for small $r_{ij}r_{jk}r_{ki}$ or
when all three angles are $\leq \frac\pi2$, and vanishes when $r_{ij}r_{jk}r_{ki}\to\infty$.
 This result guarantees clustering minimizers of $W$ for each $N\geq 3$. 

\subsection{Monotonicity of the optimal average $K$-body energy}
\label{sec:monotonicity}

 In Appendix A we establish a family of hypergraph-theoretical identities (\ref{eq:AveVkID}).
 For the $N$-body cluster models discussed below, this formalism yields, when $N>K\geq 2$, that the
average $K$-body interaction of an $N$-body cluster configuration ${\cal C}^{(N)}$ 
can be written as the arithmetic mean over the $N$ monomers, labeled by ${}_\ell$, of the average $K$-body interaction of 
the pertinent $N-1$-body cluster configurations ${\cal C}^{(N)}\backslash\{S_\ell\}$, 
\begin{equation}\label{eq:AveVidK}
\langle V\rangle({\cal C}^{(N)}) = \frac1N \sum_{1\leq \ell\leq N} \langle V\rangle({\cal C}^{(N)}\backslash\{S_\ell\});
\vspace{-.3truecm}
\end{equation}
when $K=2$ this is identity (9) in \cite{KieJCP}.

 Using (\ref{eq:AveVidK}), the reasoning of \cite{KieJCP} for $K=2$ generalizes
essentially unchanged to integers $K\geq 2$.
 Specifically, since identity (\ref{eq:AveVidK}) holds for all cluster configurations $\cC^{(N)}$, 
it holds in particular for the global energy-minimizing configuration $\cC^{(N)}_{\rm min}$. 
 But then, since each $\cC^{(N)}_{\rm min}\backslash\{S_\ell\}$ is an $N-1$-monomer configuration that is not
necessarily an energy-minimizing $N-1$-monomer configuration, by replacing each $\cC^{(N)}_{\rm min}\backslash\{S_\ell\}$ 
with the global minimum $N-1$-monomer configuration $\cC^{(N-1)}_{\rm min}$, for all $N>K$ we obtain  \vspace{-.5truecm}
\begin{equation}
\langle V\rangle \big(\cC^{(N)}_{\rm min}\big) = 
\frac1N \sum_{\ell=1}^N  \langle V\rangle \big(\cC^{(N)}_{\rm min}\backslash\{S_\ell\}\big) 
\geq \frac1N \sum_{\ell =1}^N  \langle V\rangle \big(\cC^{(N-1)}_{\rm min}\big) 
= \langle V\rangle \big(\cC^{(N-1)}_{\rm min}\big)
\vspace{-.5truecm}
\end{equation}
(cf.~\cite{KieJCP} for when monomers interact solely pairwise).
  In short, for all $N\geq K$,
\begin{equation}\label{vMONOlaw}
\centerline{\boxed{\phantom{\Big[}v(N+1) \geq v(N) \phantom{\Big]}}.}\hspace{-1truecm}
\end{equation}
  This a-priori inequality is satisfied by any list $N\mapsto v(N)$ of average $K$-body energies for the
global minima of $N$-monomer clusters formed with a single type of $K$-body interaction.

\subsection{Inequality (\ref{vMONOlaw}) as a necessary condition for optimality}

To keep this article self-contained (for the convenience of the reader), we include 
this subsection from \cite{KieJCP} largely verbatim.

 Since there is no known algorithm that finds the global minimum of an $N$-monomer cluster configuration in polynomial time,
even sophisticated modern day computer experiments may fail to locate it, and instead return a local minimum.
 Testing computer-generated lists of putative global minima for whether they are compatible with (\ref{vMONOlaw})
may reveal $N$-body cluster energies that are not optimal.

To fail this monotonicity test means the following. Suppose in some computer-experimentally determined
list of putatively optimal $N$-body configurations one finds for a certain $N_*$ that 
$\langle V\rangle \big(\cC^{(N_*-1)}\big) > \langle V\rangle \big(\cC^{(N_*)}\big)$. 
 In that case, and provided that there is no error in the transcription of the data,
 one can conclude that the $N_*-1$-monomer cluster is certainly not optimal.
 As a consequence of the $N_*-1$-monomer cluster failing the monotonicity test against the $N_*$ cluster,
even if the $N_*-2$-monomer cluster passed the test against the $N_*-1$ cluster, i.e. if
$\langle V\rangle \big(\cC^{(N_*-2)}\big) \leq \langle V\rangle \big(\cC^{(N_*-1)}\big)$, 
the $N_*-2$ cluster may still fail the monotonicity test against the $N_*$ cluster. 
 More generally, each $N_*-n$-monomer configuration for which 
$\langle V\rangle \big(\cC^{(N_*-n)}\big) > \langle V\rangle \big(\cC^{(N_*)}\big)$, with $n\geq 1$,
is not optimal.

 Note that for each non-optimal configuration with $N = N_*-n$, 
the difference $\langle V\rangle \big(\cC^{(N_*-n)}\big) - \langle V\rangle \big(\cC^{(N_*)}\big)$
is a lower estimate for the amount by which the average $K$-body energy of $\cC^{(N_*-n)}$ overshoots the optimal value.

\subsection{Using (\ref{vMONOlaw}) to bound missing energy data from above}

This subsection also carries over from \cite{KieJCP} essentially verbatim, to provide a complete account.

The monotonicity (\ref{vMONOlaw}) of the map $N\mapsto v(N)$ 
for optimal average pair energies implies that any value $\langle V\rangle \big(\cC^{(N)}\big)$ 
computed with a putatively optimal $\cC^{(N)}$ is an upper bound on all $v(\tilde{N})$ with $\tilde{N}<N$.
 Hence, even if no putatively optimizing configurations $\cC^{(\tilde{N})}$ have yet been computed for certain $\tilde{N}<N$,
 (\ref{vMONOlaw}) yields some information about the pertinent missing optimal energies 
in such lists of putative global minimum energies that have gaps.

\subsection{On the tightness of the monotonicity law (\ref{vMONOlaw})}\label{sec:tightness}

 The monotonicity law (\ref{vMONOlaw}) furnishes a necessary but not sufficient condition for optimality. 
 Thus, a list of putatively optimal configurations that satisfies the monotonicity
law (\ref{vMONOlaw}) does not necessarily feature true global minima configurations. 
 The more likely a sub-optimal minimum is to fail the test, the ``tighter'' and more useful the test is.
Here we consider how many local minima may pass the test, in addition to the true global minimum.
 This is a difficult question for which we have obtained some preliminary empirical insight for the special
but important $K=2$ case of Lennard-Jones clusters, where some relatively large databases have been
harvested at particular sizes {$N$} to address global thermodynamic and kinetic properties \cite{Wales13}.
For these examples, only the set of local minima for $N=13$ is likely to be complete.
For the other sizes, the sampling is most extensive for the low-lying local {energy} minima, which enables 
us to provide a lower bound on the number of local {minimizers} (each simply denoted ${\cal C}^{(N)}_{\rm loc}$) 
that satisfy the inequality
\begin{equation}\label{UPPERboundWofN}
\langle V\rangle \big(\cC^{(N)}_{\rm loc}\big)
\leq 
v(N+1).
\end{equation}
 Note that even if $v(N+1)$ is not rigorously known, we may replace it by $v^x(N+1)$, its best upper approximation
found in the list of computer-experimental locally optimal configurations.
 The result surprised us: already for moderately large $N$ there are many non-optimal local minimizers that satisfy 
(\ref{UPPERboundWofN}), as shown in Table \ref{table:local}, 
{where we list the absolute numbers of minima that pass the monotonicity test for each $N$.} 
 
\begin{table}[H]
\begin{center}
\begin{tabular}{||c c c r||} 
 \hline
 $N$ & $v^x_{\mbox{\tiny{LJ}}}(N)$ & $v^x_{\mbox{\tiny{LJ}}}(N+1)$ & 
 $\#\{\langle V_{\mbox{\tiny{LJ}}}\rangle \big(\cC^{(N)}_{\rm loc}\big)\leq v^x_{\mbox{\tiny{LJ}}}(N+1)\}$ \\ 
 \hline\hline
 13 & $-$0.5682923205 & $-$0.5257709487 & 4 $\phantom{x(N+1)}$\\ 
 \hline
  31 & $-$0.2872826280 & $-$0.2815232341 & 4,331 $\phantom{x(N+1)}$\\
 \hline
 38 & $-$0.2474088578 & $-$0.2429597624 & 395 $\phantom{x(N+1)}$\\ 
 \hline
 55 & $-$0.1880461077 & $-$0.1841838343 & 349 $\phantom{x(N+1)}$\\ 
 \hline
  75 & $-$0.1432404796 & $-$0.1413666195 & 74,030 $\phantom{x(N+1)}$\\ 
 \hline
\end{tabular}
\caption{\label{table:local} The number of local minima that satisfy the monotonicity test in
databases of Lennard-Jones clusters obtained in previous work. These values are lower bounds, aside from $N=13$.
The greater values for $N=31$ and $N=75$ result from more extensive sampling compared to
$N=38$ and $N=55$, since larger databases were created to converge thermodynamic properties in previous work \cite{Wales13}.}
\end{center}
\end{table}
\vspace{-.75truecm}

Table \ref{table:local} includes results for two of the magic sizes where complete Mackay icosahedra are possible,
{i.e. $N=13$ and $55$}.
In contrast, the landscapes for $N=31$, $38$, and $75$
are of interest because they exhibit double-funnel organisation \cite{DoyeMillerWalesA},
with competing low-energy structures separated by high barriers.
The results in Table \ref{table:local} partly reflect much more extensive sampling for sizes
$N=31$ and $N=75$, where we aimed to converge the calculated thermodynamic properties accurately \cite{Wales13}.

\begin{figure}[H]
\begin{tabular}{cc}
\includegraphics[width=0.45\textwidth]{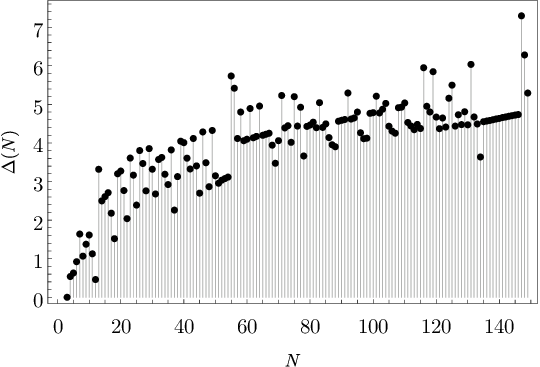} & \includegraphics[width=0.45\textwidth]{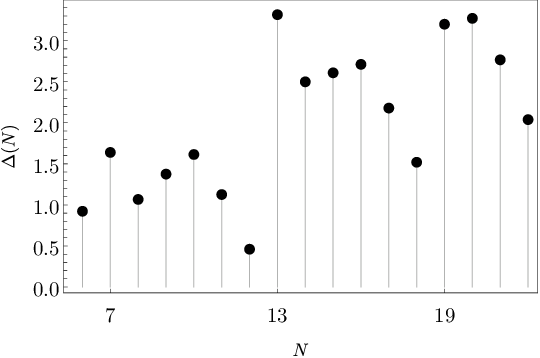} \\
\includegraphics[width=0.45\textwidth]{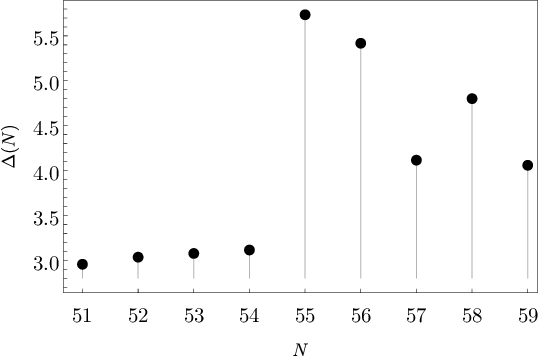} & \includegraphics[width=0.45\textwidth]{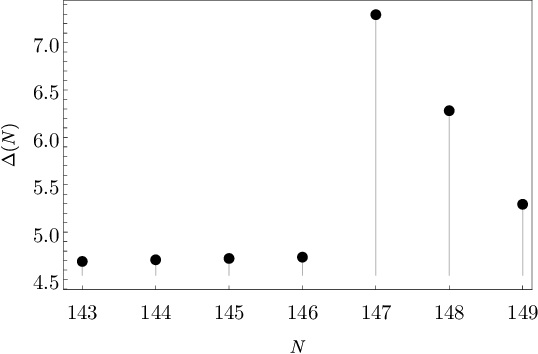}
\end{tabular}
\caption{\label{fig:Delta}
 Plots of $\Delta^x(N)$ for LJ$_N$ clusters, showing the difference between the {empirical} upper bound
imposed by the monotonicity condition {on the actual global minimum energy}
and the {empirical} global minimum energy as a function of $N$. 
{Here we qualify the data as empirical because we are using the lowest
known minima reported from existing searches.}
The three panels that follow the full plot magnify regions around the Mackay icosahedra at $N=13$, $55$ and $147$.
In each case $\Delta^x(N)$ exhibits a clear step {up}.
The pentagonal bipyramid at $N=7$ and the double icosahedron at $N=19$ also exhibit 
{clear steps up} corresponding to enhanced local stability.}
\end{figure}

 {Another factor that contributes to the large value for $N=75$ is the 
trend for $\Delta(N)$ to increase with $N$ overall, 
where $\Delta(N) := \frac{N-1}{N+1} W\big(\cC^{(N+1)}_{\rm min}\big) - W \big(\cC^{(N)}_{\rm min}\big)$
is the difference between the upper bound $\frac{N-1}{N+1}  W\big(\cC^{(N+1)}_{\rm min}\big)$
on the actual global minimum energy $W \big(\cC^{(N)}_{\rm min}\big)$ 
and the actual global minimum energy itself, as obtained from the $K=2$ monotonicity test.}
  This trend results from the increasing average coordination number for these clusters, where many atoms
lie in a surface environment.
 In Figure \ref{fig:Delta} we plot {$\Delta^x(N)$, the computer-experimental counterpart of $\Delta(N)$,} 
for LJ$_N$, including magnifications of the
regions containing magic number clusters based on Mackay icosahedra \cite{mackay62} at $N=13$, $55$ and $147$.
 These sizes correspond to particularly favourable geometric packings with special stability \cite{WalesBOOK}, and we see that
{in each case there is a significant step up in $\Delta^x(N)$} followed by a progressive 
{overall} decrease {towards the trend}.
 The {overall trend for $\Delta^x(N)$} to increase with $N$, {and the large step-ups,
are} evident in Figure \ref{fig:Delta}.

\vspace{-.25truecm}

{Figure \ref{fig:Delta} exhibits another interesting feature: each value of $\Delta^x(N)$ appears
to lie in the vicinity of one of several distinct concave sequences $N\mapsto D_i(N)$, $i\in\{1,2,...\}$, with bifurcations among them.
 This pattern probably reflects competition between alternative structural families.
In future work it would be interesting to connect such features with previous analysis of how structural competition reflects geometrical
effects, which can be convoluted with entropic contributions and quantum behaviour \cite{doyew96a,DoyeC02,CalvoDW01b}.}

{Judged by the absolute number counts, the monotonicity test appears not to be 
very tight already when $N=31$.
 A more informative assessment would require accurate relative frequencies.
 Unfortunately, so far no meaningful relative frequencies can be computed for the displayed $N$ values, with the exception of $N=13$.
 Despite having millions of local LJ$_N$ minima in the available databases, we only have a small fraction of the total.
From the potential energy density of local minima calculated in \cite{Wales13}
we can see that the total number 
will be over $e^{35}$ and $e^{50}$ for LJ$_{31}$ and LJ$_{75}$, respectively. 
 This of course is a manifestation of the NP-hardness of the problem, for which more
generally the number of local minimizers has been estimated (non-rigorously) to grow exponentially with $N$, see 
\cite{StillWebA}, \cite{StillWebB}, \cite{DoyeWales2002}, \cite{WalesDoye2003}.}
 {In this vein, we wonder whether more exhaustive studies of the energy landscape will reveal 
that the number of local minimizers passing the monotonicity test also grows exponentially with $N$.
 Extensive empirical studies, combined with more rigorous analysis, will be needed to test this hypothesis.}

{It certainly is desirable to tighten the bound (\ref{vMONOlaw}) somehow,
 but it does not appear to be straightforward.}
 Recalling the definition (\ref{optVKave}), and the ensuing equivalent version (\ref{optVKaveREWRITE}), of the
optimal average $K$-body energy of a pertinent $N$-body cluster, we see that the monotonicity law (\ref{vMONOlaw}) states that
\begin{equation}\label{monoLAWexplicit}
\genfrac{}{}{0.5pt}{0}{{}_1^{}}{\genfrac{(}{)}{0pt}{1}{N+1}{K}} W({\cal C}^{(N+1)}_{\rm min})
\geq 
\genfrac{}{}{0.5pt}{0}{{}_1^{}}{\genfrac{(}{)}{0pt}{1}{N}{K}} W({\cal C}^{(N)}_{\rm min}).
\end{equation}
 Multiplying (\ref{monoLAWexplicit}) by $\genfrac{(}{)}{0pt}{1}{N}{K}$,
then cancelling many terms in the resulting ratio of the two combinatorial brackets, and after a minor algebraic rewriting, one obtains
an upper bound for $W({\cal C}^{(N)}_{\rm min})$ in terms of $W({\cal C}^{(N+1)}_{\rm min})$, $N$, and $K$, viz.
\begin{equation}\label{WupperBOUND}
W({\cal C}^{(N)}_{\rm min})
\leq 
\left(1-\tfrac{K}{N+1}\right) W({\cal C}^{(N+1)}_{\rm min}).
\end{equation}
Note that $0 < 1-\frac{K}{N+1} <1$ because $0<K\leq N$.
Since $1 + {\cal O}\big(\frac1N\big) \approx 1$ for large $N$, 
one may be tempted to speculate that the ${\cal O}\big(\frac1N\big)$ term in the parenthetical factor at r.h.s.(\ref{WupperBOUND})
can be neglected.
(With $K$ fixed and typically small, say $K=2$ or $K=3$, for $N > 100$ one has less than 3$\%$ relative error in the upper
bound of $W({\cal C}^{(N)}_{\rm min})$.)
 However, this conclusion would be \emph{false} at least for all known chemical models of clusters in nature
--- in particular, those in \S \ref{sec:Baxter}, including the special case of the Lennard-Jones clusters, 
and all the familiar variations on their theme for which $W({\cal C}^{(N)}_{\rm min}) <0$ and 
\begin{equation}
W({\cal C}^{(N)}_{\rm min})
\geq 
W({\cal C}^{(N+1)}_{\rm min}).
\end{equation}
 Thus the term $\frac{K}{N+1}$, though tiny for large $N$, is decisive for the inequality (\ref{WupperBOUND}) to be true. 
 Hence there does not seem to be much room to improve on (\ref{vMONOlaw}).

 This result does not mean there are no avenues left to explore.
 For instance, the equivalent inequality (\ref{WupperBOUND}) is the special case $p=1$ of the family of inequalities
\begin{equation}\label{WupperBOUNDp}
W({\cal C}^{(N)}_{\rm min})
\leq 
\left(1-\left[\tfrac{K}{N+1}\right]^p\right) W({\cal C}^{(N+1)}_{\rm min}); \ p \in [1,p^*],
\end{equation}
where $p^*$ is the largest $p$-value for which (\ref{WupperBOUNDp}) is true for all $N\geq K$.
 When $p=1$ the inequality in (\ref{WupperBOUNDp}) is true, but when $p\to\infty$ it
is certainly false for cluster models of interest in chemical physics, 
 where $1\leq p^*<\infty$, as explained above.
 The problem is to find $p^*$, which may of course depend on $K$.
 If $p^*>1$, then (\ref{WupperBOUNDp}) with $p=p^*$ furnishes a tighter test than our (\ref{vMONOlaw}), yet if
$p^*=1$ then our monotonicity law (\ref{vMONOlaw}) is optimal within the family of laws of the type (\ref{WupperBOUNDp}).

 Determining $p^*$ remains a problem for future work. 

\section{On optimal clustering for additive mixtures of irreducible $k$-body potentials with different $k$}
\label{sec:MULTIk}
\vspace{-.15truecm}

For this extension the definitions and assumptions of \S \ref{sec:K} need only minor adjustments.

 We maintain the notation for the \emph{state} of a monomer, denoted by $S\in {\mathfrak{S}}$, with 
 $\mathfrak{S}$ the \emph{monomer state space}.
Also ${\cal C}^{(N)} :=\{ S_1, ..., S_N\}$ will continue to denote the \emph{configuration} of an $N$-monomer cluster,
with no two of the $N$ state variables in ${\cal C}^{(N)}$ coinciding.
 
 Since we here allow the monomers in a cluster to interact with an additive mix of various different $k$-body 
potentials, we add a suffix ${}_k$ to $V$.
 Thus, for a fixed integer $K\geq 2$ 
and $k\in \{1,...,K\}$, by $V_k$ we denote a real-valued, permutation-symmetric $k$-state potential energy.
 For the sake of notational convenience we allow the
degenerate case $V_k\equiv 0$ for some $k<K$, yet $V_K$ will always be irreducible, hence non-trivial.
 In particular, setting $V_k\equiv 0$ for all $k<K$ reduces this section to \S \ref{sec:K}.
 Whenever $V_k\not\equiv 0$ we again stipulate it to be irreducible.
 Here, ``permutation-symmetric'' and ``irreducible'' have the same meanings as in \S \ref{sec:K}; 
note that permutation invariance and irreducibility are nontrivial conditions on $V_k$ only when $k>1$.

 For  $N\geq K$, with $K\geq 2$, the total potential energy of an $N$-monomer cluster considered here is now given by
\begin{equation}\label{WdefMULTIk}
W({\cal C}^{(N)} ) 
= 
\sum_{1\leq i \leq N} V_1(S_i)
+
\sum\!\!\sum_{\hskip-.6truecm 1\leq i<j \leq N} V_2(S_i,S_j)
+
\cdots
+
\sum\; \cdots\!\!\!\sum_{\hskip-1.2truecm 1\leq i_1<\cdots<i_K \leq N} V_K(S_{i_1},...,S_{i_K}).
\end{equation}
 In analogy to \S \ref{sec:K} it is assumed that for each $N\geq K\geq 2$
there exists a globally minimizing $N$-monomer configuration ${\cal C}^{(N)}_{\rm min}$ of $W$ defined in (\ref{WdefMULTIk});
note that $K\geq 2$ is still needed for cluster formation.
 Note also that, in contrast to \S \ref{sec:K}, the expression (\ref{WdefMULTIk}) for $W$ now may include internal single-monomer 
energy $V_1$ (accounting for internal distortions such as bending, twisting and stretching).

 Next we define the average $k$-body energy of an $N$-monomer configuration by
\begin{equation}\label{eq:AveVk}
\langle V_k\rangle({\cal C}^{(N)}) := 
\genfrac{}{}{0.5pt}{0}{{}_1^{}}{\genfrac{(}{)}{0pt}{1}{N}{k}}
\sum\; \cdots\!\!\!\sum_{\hskip-1.2truecm 1\leq i_1<\cdots<i_k \leq N} V_k(S_{i_1},...,S_{i_k}); \quad k\in\{1,...,K\}.
\end{equation}
 In terms of (\ref{eq:AveVk}), the total energy (\ref{WdefMULTIk}) of a configuration ${\cal C}^{(N)}$ 
can be recast as
\begin{equation}\label{WofMULTIkAVEsum}
W({\cal C}^{(N)} ) 
= 
\sum_{1\leq k\leq K} 
{\genfrac{(}{)}{0pt}{1}{N}{k}}
\langle V_k\rangle({\cal C}^{(N)}).
\end{equation}
 In this format it is clear that when irreducible $k$-body potentials with different $k$ are
involved, then typically none of the average $k$-body energies is optimized by the optimizer 
$\cC^{(N)}_{\rm min}$ of $W$.

This ends our list of general definitions and assumptions for this section.
 When only a single $k=K$ interaction is involved in the model (in which case $V_1\equiv 0$), these assumptions
are all that is needed to establish the monotonicity result (\ref{vMONOlaw}). 
 However, when at least one $V_k\not \equiv 0$ with $k<K$, then additional cluster model assumptions are needed.
 We next will establish a family of monotonicity results for cluster global minima, valid only for certain models.

\smallskip

\subsection{\hspace{-5pt}
Monotonicity when subcluster energies are non-positive for $k < \kappa$ and non-negative for $k>\kappa$, with $1<\kappa\leq K$}

Our many-$k$ monotonicity result is valid under the following additional assumption. 

(A): There is a fixed $\kappa\in\{2,...,K\}$ such that for all $N\geq K+1$, all nontrivial 

\qquad \ $k$-body energies of each $N-1$-body subcluster of the $N$-body ground state 

\qquad \ $\cC^{(N)}_{\rm min}$ are non-positive if $k<\kappa$ and non-negative if $k > \kappa$.

 Notice that assumption (A) leaves open the sign of the $\kappa$-body energies. 
 Notice furthermore that identically vanishing interactions are both non-negative and non-positive.
 In particular, when $\kappa = K$, then the $k$-body energies with $k>\kappa$ simply vanish, compatible
with assumption (A).

Assumption (A) in concert with our general assumptions yields the following monotonicity result.
Define 
\begin{equation}\label{wDEF}
w(N) : = \genfrac{}{}{0.5pt}{0}{{}_1^{}}{\genfrac{(}{)}{0pt}{1}{N}{\kappa}} W\big(\cC^{(N)}_{\rm min}\big).
\end{equation}
 Then for all $N>K$ we have
\begin{equation}\label{wMONOlaw}
\centerline{\boxed{\phantom{\Big[}w(N+1) \geq w(N) \phantom{\Big]}}.}\hspace{-1truecm}
\end{equation}

\emph{Proof of} (\ref{wMONOlaw}):
By the hypergraph-theoretical identity (\ref{eq:WidANYk}) we have
\begin{equation}\label{eq:Wid}
W\big(\cC^{(N)}_{\rm min}\big) 
=
\sum_{1\leq \ell\leq N}\sum_{1\leq k\leq K}
\frac{1}{N-k}
\sum\; \cdots\!\!\sum_{\hskip-1.3truecm \begin{subarray}{c} 1\leq i_1<\cdots<i_k \leq N 
\\ \!\! \ell\not\in \{i_1,...,i_k\}       \end{subarray}                      } 
V_k(S_{i_1},...,S_{i_k}).
\end{equation}
 We now multiply (\ref{eq:Wid}) with $N-\kappa$. 
Since $k\leq K<N$, for $k<\kappa$ we have $0<\frac{N-\kappa}{N-k} < 1$, and for $\kappa<k\leq K$ we have
$1 <\frac{N-\kappa}{N-k} < \infty$.
 Thus, assumption (A) now implies that 
\begin{equation}\label{eq:WineqA}
(N-\kappa)W\big(\cC^{(N)}_{\rm min}\big) 
\geq
\sum_{1\leq \ell\leq N}
\sum_{1\leq k\leq K}\ 
\sum\; \cdots\!\!\sum_{\hskip-1.3truecm \begin{subarray}{c} 1\leq i_1<\cdots<i_k \leq N 
\\ \!\! \ell\not\in \{i_1,...,i_k\}       \end{subarray}                      } 
V_k(S_{i_1},...,S_{i_k})
= 
\sum_{1\leq \ell\leq N}
W\big(\cC^{(N)}_{\rm min}\backslash\{S_\ell\}\big).\vspace{-1truecm}
\end{equation}
 We now can argue as in the proof of (\ref{vMONOlaw}).
 Since each $\cC^{(N)}_{\rm min}\backslash\{S_\ell\}$ is an $N-1$-monomer configuration that is not
necessarily an energy-minimizing $N-1$-monomer configuration, by replacing each $\cC^{(N)}_{\rm min}\backslash\{S_\ell\}$ 
with the $\ell$-independent energy-minimizing $N-1$-monomer configuration $\cC^{(N-1)}_{\rm min}$, for all $N>K$ we obtain  
\begin{equation}\label{eq:WineqB}
(N-\kappa)W\big(\cC^{(N)}_{\rm min}\big) 
\geq
N W\big(\cC^{(N-1)}_{\rm min}\big).
\end{equation}
Dividing left- and right-hand sides of (\ref{eq:WineqB}) by $(N-\kappa){\genfrac{(}{)}{0pt}{1}{N}{\kappa}}$ 
yields (\ref{wMONOlaw}). \hfill\emph{End of proof}.
\smallskip

We now list some examples. 

\vspace{-.3truecm}

\subsection{Examples of models with multiple $k$-body interactions}
\label{sec:multiple}

\subsubsection{Optimal distancing (and related issues) on a sphere}
\label{sec:sphere}

Our first example is a generalization of the problem to maximize the arithmetic mean of the Euclidean distance 
between $N$ points on the sphere $\Sset^2$.
\smallskip

We take $\mathfrak{S} = \Sset^2\subset\Rset^{3}$, so
$S = \rV\in\Rset^{3}\cap\{|\rV| =1 \}$ is a unit-length position vector
in three-dimensional Euclidean space, representing the location of a point particle on the unit sphere in $\Rset^3$.
 For $N\geq 4=K$, the $N$-body cluster energy is taken to be 
\begin{equation}\label{Wdistancing}
\!\!\!
W({\cal C}^{(N)} ) 
= 
\textstyle{\sum\!\!\sum\limits_{\hskip-.6truecm 1\leq i<j \leq N}} V_2(\rV_i,\rV_j)
+
\textstyle{\sum\!\sum\!\!\sum\limits_{\hskip-1truecm 1\leq i<j<k \leq N}} V_3(\rV_{i},\rV_{j},\rV_{k})
+
\textstyle{\sum \cdots\!\!\!\sum\limits_{\hskip-1.1truecm 1\leq i_1<\cdots<i_4 \leq N}} V_4(\rV_{i_1},...,\rV_{i_4}),
\end{equation}
where 
\begin{equation}
V_2(\rV_i,\rV_j)= - |\rV_i-\rV_j|
\end{equation} 
is the negative of the Euclidean distance of points $i$ and $j$, 
\begin{equation}
V_3(\rV_i,\rV_j,\rV_k)
 := - \alpha \left| (\rV_k-\rV_j)\boldsymbol{\times}(\rV_i-\rV_j) \right|
\end{equation}
with $\alpha \geq 0$ is a non-positive multiple of the area of the parallelogram in $\Rset^3$ 
spanned by $\rV_i-\rV_j$ and $\rV_k-\rV_j$, while
\begin{equation}
V_4(\rV_{i_1},\cdots,\rV_{i_4})
 := 
-\beta \left|\bigl( (\rV_{i_2}-\rV_{i_1})\boldsymbol{\times}(\rV_{i_3}-\rV_{i_1}) \bigr)\cdot (\rV_{i_4}-\rV_{i_1}) \right|
\end{equation}
with $\beta\in\Rset$ is a real multiple of the volume of the parallel-epiped spanned by the three $\rV_{i_\ell}-\rV_{i_1}$, 
$\ell\in\{2,3,4\}$.
 In this example all nontrivial $k$-body interactions with $k\leq 3$ are $\leq 0$ for any $k$-body subset, while 
those with $k=4$ may be non-positive, non-negative, or vanish identically. 
 Hence, for $\beta>0$ assumption (A) is satisfied with $\kappa=4$; for $\beta=0< \alpha$ it is satisfied with $\kappa=3$; 
when $\beta<0$ assumption (A) is also satisfied with $\kappa=3$.

Since in the special case $\alpha=0=\beta$ the model reduces to the well-known problem of maximizing the arithmetic 
mean distance among $N$ points on the sphere $\Sset^2$ \cite{Beck}, when
$\alpha$ or $\beta$ are non-zero but small in magnitude, the problem may thus
 be treated as a small perturbation of the optimal distancing problem on $\Sset^2$.

 The non-perturbative regime is much more complicated, and more interesting. 
 For instance, setting $\alpha =0$ and $\beta<0$ with $|\beta|\gg 1$ the volume term penalizes any configuration whose 
points are not all on the equator $\Sset^1\subset\Sset^2$. 
 In the limit $\beta\to -\infty$ the optimizers are forced to be concentrated on $\Sset^1$, thus are
$N$-point configurations that maximize the arithmetic mean distance on $\Sset^1$, and these are all known, 
namely regular $N$-gons. 
 The interesting open question is how for $\beta$ large negative, but not too large, the competition between the distance and
the volume terms will play out. 

Final remark: 
The three-point and four-point interaction terms, here invoked as geometrically 
natural many-body analogues of the prominent two-point interaction term, 
feature on their own in the Heilbronn problem (cf. \S \ref{sec:trimer}), respectively in a three-dimensional analog of the Heilbronn 
problem \cite{Lefmann}.

\medskip

\subsubsection{Clustering with induced dipole-dipole interactions}
\label{sec:LJATG} 

Our second example is a $K=2$ model motivated by the fact that the attractive $-1/r^6$ term of the Lennard-Jones 
potential is the result of an induced dipole-dipole interaction of polarizable monomers.
 This interaction is implemented here in a toy version, which employs the interplay of a dipole-dipole with a single-monomer potential.

We take $\mathfrak{S} = \Rset^3\times\Sset^2\times [0,1]$, with $S=(\rV,\bomega,\wp)$, 
where $\rV$ represents the position of a polarizable monomer, $\wp$ the strength of
its polar moment, and $\boldsymbol{\omega}$ is a unit vector representing the orientation of the polar moment. 
  For $N\geq 2$, the $N$-body cluster energy is taken to be 
\begin{equation}\label{WforHCwithDIP}
W\big({\cal C}^{(N)} \big) 
= 
\sum_{1\leq i \leq N} V_1(S_i)
+
\sum\!\!\sum_{\hskip-.6truecm 1\leq i<j \leq N} V_2(S_i,S_j),
\end{equation}
where the single-monomer potential is chosen to be
\begin{equation}\label{Vone}
V_1(S) := -1 + \wp^4,
\end{equation}
and the pair potential 
$V_2(S_i,S_j) := V_{\mbox{\tiny{hs}}}(|\rV_i-\rV_j|) 
+  V_{\mbox{\tiny{dip}}}^{\mbox{\tiny{tr}}}(|\rV_i-\rV_j|;\bomega_i\cdot\bomega_j;\wp_i\wp_j)$,
with
\begin{equation}\label{Vhc}
V_{\mbox{\tiny{hs}}}(r) 
:= \begin{cases}\, \infty \quad \mbox{if} \quad r < 1 \\ \ 0 \quad\ \mbox{if} \quad r \geq 1
\end{cases}
\end{equation}
the (non-sticky) hard sphere interaction for spheres of radius $\frac12$ and 
\begin{equation}\label{VLJpolar}
V_{\mbox{\tiny{dip}}}^{\mbox{\tiny{tr}}}(r;\cos\gamma;\lambda) 
:= - 4\alpha \lambda\frac{\cos\gamma}{r^3}
\end{equation}
a truncated dipole-dipole interaction [cf. (\ref{Vdip})].
The coupling constant $\alpha \in \Rset$ is a small parameter (but considered fixed).
 To have clustering we need $\alpha\neq 0$.
 Yet it is instructive to first discuss the non-clustering case $\alpha=0$, which
gives a system of $N$ polarizable (but not sticky) hard spheres.
\medskip

\noindent
\textbf{3.2.2.a} \emph{The case $\alpha=0$.}
 The energy functional (\ref{WforHCwithDIP}) in this case has a minimum $-N$, which is achieved by \emph{any} 
configuration $\cC^{(N)}_{\rm{min}}$ of non-polarized monomers with pairwise distances $|\rV_i-\rV_k|\geq 1$ for $i\neq j$;
note that the single-monomer potential takes its minimum value $-1$ when $\wp=0$.
 Thus the general model assumptions and assumption (A) are satisfied with $\kappa=2=K$.
 Yet the vast majority of local minimum configurations do not qualify as clusters, and those that have the appearance of a 
cluster appear so only incidentally.

The set of all such configurations of $N$ non-polarized hard spheres is a ``critical plateau region'' of the 
energy functional (\ref{WforHCwithDIP}) when $\alpha\neq 0$, but none of these configurations is then energy-minimizing.
 Instead, the energy-minimizing $N$-monomer configurations when $\alpha\neq 0$ are genuine $N$-body clusters, shown
next.
\medskip

\noindent
\textbf{3.2.2.b} \emph{The case $\alpha\neq 0$.} 
 Assumption (A) is manifestly satisfied with $\kappa=2$ for any minimizer $\cC^{(N)}_{\rm min}$ --- provided that
the infimum of $W\big({\cal C}^{(N)} \big)$ is in fact achieved by some ${\cal C}^{(N)}_{\rm min}$.
 We now show that this is the case, and that the minima are genuine clusters.

 The dimer problem $N=2$ deserves special attention, for it
reveals the relation of this model to the Lennard-Jones cluster problem. 
 Namely, with $\bomega_1= {\mathrm{sign}}(\alpha)\bomega_2 :=\bomega\in \Sset^2$ chosen freely, and with $\wp_1=\wp_2 :=\wp>0$ 
to be determined,
for given $\rV_1$ and $\rV_2$ satisfying $|\rV_1-\rV_2|:= r \geq 1$ the minimization of the energy functional yields
$\wp^2 = {|\alpha|}\frac{1}{r^3}$, which for the conditionally minimal energy (conditioned on $|\rV_1-\rV_2|:= r \geq 1$)
gives $W\big(\cC^{(2)}\big) = -2 - 2{\alpha^2}\frac{1}{r^6}$.
 Except for the trivial shift by $-2$ this model captures the attractive long range behavior of the Lennard-Jones 
pair potential; its repulsive short range term is replaced by the restriction that $r \geq 1$. 

The situation is more subtle when $N\geq 3$. We distinguish $\alpha>0$ and $\alpha <0$.

\noindent
\textbf{3.2.2.b.i} \emph{The case $\alpha > 0$.} 
 $N$ non-overlapping spheres of radius $\frac12$ can always be placed inside a spherical domain of
sufficiently large radius $R$, say $R=N/2$. 
 Hence the maximal distance between the position vectors of any two spheres is bounded by $N$.
 Take any such $N$-sphere configuration and set all the polar moment vectors parallel, so
$\bomega_i\cdot\bomega_j =1$ for all $1\leq i<j\leq N$, and set all moment strengths equal, viz. $\wp_i=\wp$.
 Then for such a trial configuration the energy is bounded above by
$W\big({\cal C}^{(N)}_{\rm trial} \big) \leq - N + N \wp^4 -   2\alpha\frac{N-1}{N^2}\wp^2$,
which is minimized by $\wp_*^2 =  \frac{N-1}{N^3}\alpha$, yielding the upper bound
$\inf W\big({\cal C}^{(N)}\big) \leq - N \left(1+ \frac{(N-1)^2}{N^6}\alpha^2\right)$.
 Thus we have $\inf W\big({\cal C}^{(N)} \big) < - N$, which establishes that no configuration of all 
unpolarized hard spheres can be a $N$-body minimizer when $\alpha >0$.
 Moreover, by a straightforward variation of this argument we can relax the conditions that all $\wp_i=\wp$
and that all $\bomega_i\cdot\bomega_j=1$, and prove that a 
minimizing configuration must have all its $\wp_i>0$, and all its $\bomega_i=\bomega$, with $\bomega$ any fixed unit vector.
 As to clustering, consider {any} admissible configuration with 
all $|\rV_i-\rV_k|\geq 1$ for $i\neq j$, all $\bomega_i=\bomega$, and all $\wp_i>0$ fixed, and now 
replace all $\rV_i$ by $\lambda\rV_i$, with $\lambda >1$. 
 This construction yields another admissible configuration with the same $\bomega_i$ and $\wp_i$, but all 
pair interactions (which are negative) have increased, while all single monomer contributions have remained the same. 
 Thus the interactions favor clustering, i.e.~it is only necessary to ask whether a minimizing $N$-body configuration
exists in $B_{N/2}(0)$, the closed spherical domain of radius $N/2$ centered at the origin.
 The answer is positive because $W\big(\cC^{(N)}\big)$ is continuous on the pertinent 
set of admissible configurations, which form the bounded and closed subset 
$\{\cC^{(N)}: |\rV_i-\rV_k|\geq 1\ \mbox{for}\ i\neq j,\ \mbox{and}\ 
\rV_i\in B_{N/2}(0)\ \mbox{and}\ \wp_i \in[0,1]\ \mbox{and}\ \bomega_i=\bomega\ \mbox{for\ all}\ i\}$ of configuration space.
 This derivation establishes the existence of a minimizing cluster configuration ${\cal C}^{(N)}_{\rm min}$ for each $N\geq 2$. 

 In general, the minimizing configurations are not easy to determine, but our preceding arguments demonstrate that
a minimizing configuration is a cluster with all $\bomega_i=\bomega$ and all $\wp_i>0$, and of course all 
$|\rV_i-\rV_k|\geq 1\ \mbox{for}\ i\neq j$.
 We suspect that the minimizer $\cC^{(N)}_{\rm min}$ is always a hard sphere configuration with the maximal number 
$c(N,3)$ of contact points, and suitably chosen $\wp_i>0$.
 For $N\in\{2,3,4\}$ this suggestion is readily verified
(viz., the position vectors $\rV_i$ mark an interval for $N=2$, an equilateral triangle for $N=3$, 
and a regular tetrahedron for $N=4$), with $\wp_i = \wp(N)$ for all $i\in\{1,...,N\}$, where $\wp(N)$ 
is the minimizing $\wp$-value of the quartic $N\wp^4 - 4{\genfrac{(}{)}{0pt}{1}{N}{2}}\alpha\wp^2$ for $\wp\in[0,1]$. 
 One easily finds that $\wp(N)^2 = {(N-1)}\alpha$ (here we need $\alpha \leq \frac13$), which gives for the minimum energy
 $W\big({\cal C}^{(N)}_{\rm min}\big) = - N ( 1 + (N-1)^2\alpha^2)$, with $N\in\{2,3,4\}$.
 The case $N=5$, while also counted among the ``trivial'' cases of the optimal sticky hard sphere problem (see 
\S \ref{sec:Baxter}),
is already much more complicated here because of the additional degree of freedom involving the $\wp_i$.

\smallskip

\noindent
\textbf{3.2.2.b.ii} \emph{The case $\alpha < 0$.}
 With $\alpha<0$ the arrangement $\bomega_i\cdot\bomega_j=-1$ is impossible to achieve for all pairs $i<j$ when $N>2$.
 In other words, the $\alpha <0$ problem exhibits the phenomenon of \emph{frustration}, 
where no given set of $N>2$ pairwise distinct positions of the monomers
can energetically optimize all the pair interactions at the same time. 
 Examples of systems with dimeric interactions that exhibit frustration include water clusters (e.g.~\cite{MahoneyJ00}),
and Ising spins on a triangular lattice with anti-ferromagnetic nearest-neighbor coupling (e.g.~\cite{MoessnerRamirez}).

Thus the problem of determining the global minima is more challenging than when $\alpha>0$.
 The only trivial case is the optimal arrangement of $(\rV_i,\bomega_i,\wp_i), i\in\{1,2\}$ when $N=2$. 
 However, one can demonstrate the existence of a minimal energy configuration that is a cluster, with all $\wp_i>0$,
 by noting that $\inf W\big(\cC^{(N+1)}\big)< \inf W\big(\cC^{(N)}\big)< -N$ 
for all $N\geq 2$, with all $\rV_i$ in a ball of sufficiently large radius $R(N)$. 
 The existence of a minimizer then follows from the continuity of $W$ on the compact subset of configuration space 
having energy $W \big(\cC^{(N)}\big)\leq - N -\epsilon$.

\subsubsection{Clusters with Lennard-Jones plus Axilrod--Teller interactions}
\label{sec:LJAT1}

 Our next example is more tentative, in the sense that assumption (A) is \emph{presumably satisfied}, 
although this is not manifestly obvious. 
 We will argue that it is satisfied in a perturbative regime.
 This model has already been discussed in the chemical physics literature,
in terms of the favoured structures and rearrangement mechanisms \cite{wales90a,doyew92,walesmd96},
and as a benchmark for visualisation and comparison of alternative geometry 
optimisation algorithms \cite{wales93d,AsenjoSWF13}.
Some tests of monotonicity will be considered in \S \ref{sec:LJAT}.

\smallskip

 We take $\mathfrak{S} = \Rset^3$, and $S=\rV$ represents the position of the center of mass of a polarizable monomer.
 For $N\geq 3$, the $N$-body cluster energy is taken to be 
\begin{equation}\label{WforLJAT}
\!\!\!
W({\cal C}^{(N)} ) 
= 
\sum\!\!\sum_{\hskip-.6truecm 1\leq i<j \leq N} V_2(\rV_i,\rV_j)
+
\sum\!\sum\!\sum_{\hskip-1.2truecm 1\leq i<j<k \leq N} V_3(\rV_{i},\rV_{j},\rV_{k}),
\end{equation}
where $V_2(\rV_i,\rV_j) := V_{\mbox{\tiny{LJ}}}(\rV_i,\rV_j) $
is the Lennard-Jones pair interaction (\ref{eq:VLJ}), and where
$V_3(\rV_i,\rV_j,\rV_k) := Z V_{\mbox{\tiny{AT}}}(\rV_i,\rV_j,\rV_k)$ 
is a multiple of the Axilrod--Teller trimer interaction
\begin{equation}
V_{\mbox{\tiny{AT}}}
(\rV_i,\rV_j,\rV_k) :=
\frac{1+3 \cos\gamma_i \cos\gamma_j\cos\gamma_k}{r_{ij}^3r_{jk}^3r_{ki}^3}.
\end{equation}
 As in example \S\ref{sec:water}, 
$r_{jk} :=|\rV_j-\rV_k|$ is the  Euclidean distance between the position vectors of monomers 
${}_j$ and ${}_k$, but $\gamma_j\in[0,\pi]$ is defined by $(\rV_k-\rV_j)\cdot(\rV_i-\rV_j)= :r_{kj}r_{ji}\cos\gamma_j$.
 The trimeric interaction $V_{\mbox{\tiny{AT}}}$ favors alignment of the trimer when $Z>0$ (the expected sign from quantum mechanics) 
and triangulation when $Z<0$ (a hypothetical case).

 Since the Lennard-Jones pair interaction is $<0$ for $r>1$, vanishing for infinite separation,
 and since the Axilrod--Teller trimer interaction is $<0$ 
whenever one of the angles is $\approx \pi$, and $>0$ 
when all three angles are $\leq \frac\pi2$, and vanishes when $r_{ij}r_{jk}r_{ki}\to\infty$,
and since, furthermore, the Lennard-Jones pair interaction tends to $+\infty$ faster than
the Axilrod--Teller trimer interaction can go to $-\infty$ when two monomer positions of a trimer approach each 
other arbitrarily closely, the existence of a minimizer $\cC^{(N)}_{\rm min}$
of $W$ for each $N\geq 3$ is guaranteed for any real $Z$. 

 We expect that assumption (A) holds with $\kappa=3$ and sufficiently small $|Z|\ll 1$. 
 The reason is that for $Z=0$ the problem reduces to the Lennard-Jones cluster problem, 
which is expected to be very similar to the more extreme case with the $1/r^{12}$ term replaced 
by the hard spheres interaction of example \S \ref{sec:LJATG}, in which case all subclusters of $\cC^{(N)}_{\rm min}$ 
---  not only the $N-1$-body subclusters --- do have non-positive pair energy. 
 More precisely, if we replace the repulsive $1/r^{12}$ term in (\ref{eq:VLJ}) by the repulsive $1/r^p$ with $p\geq 12$,
then for each $N\geq 2$ the ground states and their energies for this $p$-family of Lennard-Jones-like cluster problems
converge to the pertinent ground states and their energies of the hard sphere model with attractive $-1/r^6$ pair
interactions if one sends $p\to\infty$.
 By continuity there is a left neighborhood of $p=\infty$ in which all subclusters of $\cC^{(N)}_{\rm min}$ do 
have non-positive pair energy for the pertinent members of the $p$-family of Lennard-Jones-like cluster problems. 
 We expect that this neighborhood contains $p=12$, but we are not aware of a proof. 
If our expectation is correct and all subclusters of $\cC^{(N)}_{\rm min}$ 
have non-positive pair energy for the two-body 
Lennard-Jones cluster problem, then by continuity this will be the case for the mixed dimeric / trimeric
$Z\neq 0$ problem, at least with small $|Z|\ll 1$, and then assumption (A) will be satisfied with $\kappa=3$.
 
\smallskip

\section{Testing lists of $N$-monomer cluster energies with Lennard-Jones plus Axilrod--Teller interactions }
\label{sec:LJAT}
\vspace{-.15truecm}

In this section we report the outcomes of our monotonicity test (\ref{wMONOlaw}) run on
databases of minima for LJAT$_N$ clusters of $N$ monomers that interact via both the pairwise
Lennard-Jones potential and the three-body Axilrod--Teller potential, i.e.~the model explained in \S \ref{sec:LJATG}.

 The data lists were generated as follows.
 We started with the putatively optimal Lennard-Jones clusters for $N\in \{3,...,150\}$, available at 
\cite{CCD}; cf. \cite{Shao} and \cite{BRa}.
 For the range of $N$ values used in our study the optimal Lennard-Jones cluster problem has been investigated
so thoroughly (see \cite{Northby}, \cite{Leary}, \cite{DoyeMillerWalesA}, \cite{GC}, \cite{LocatelliSchoenA}, \cite{LXH},
\cite{FormanCameron}, \cite{BRb}, \cite{MuellerSbalzarini},
\cite{XJCSinJCP2004}, \cite{YWCWinJCP2019}, \cite{Doye}) 
that we confidently assume the clusters listed at \cite{CCD} are the correct global LJ$_N$ minimizers. 
 We remark that all these (putatively) optimal LJ$_N$ data of \cite{CCD} satisfy (\ref{vMONOlaw}).

 We then ``switched on'' the Axilrod--Teller interaction of model \S \ref{sec:LJATG} by setting
$Z>0 $ in incremental steps and minimising using the custom limited-memory
\cite{lbfgs,Nocedal80} quasi-Newton Broyden \cite{Broyden70}, Fletcher
\cite{Fletcher70}, Goldfarb \cite{Goldfarb70}, Shanno \cite{Shanno70} (L-BFGS) routine
in the {\tt OPTIM} program \cite{OPTIM}.
 We also checked two values of $Z<0$.
With a reasonably small maximum step size (we used a maximum of 0.2 in terms of the Euclidean distance)
we expect this relaxation to produce
the local minimum for the LJAT$_N$ cluster nearest to the putatively optimal LJ$_N$ cluster.
 For small enough $|Z|$ one may treat the AT interaction perturbatively, so that the putative LJAT$_N$ minimum
can be expected to be a slight deformation of the pertinent LJ$_N$ minimum, 
and hence the relaxation algorithm should produce this putative LJAT$_N$ optimizer.
 Assuming it does, the question then becomes: How small must $|Z|$ be to be small enough?
 Our monotonicity test (\ref{wMONOlaw}) cannot answer this question, but it can yield information about
which $|Z|$ are \emph{not} small enough whenever the test is applicable.
 
 We begin by noting that the optimizers of the dimeric limiting case $Z=0$, i.e.~the optimal LJ$_N$ 
energy data, obey (\ref{wMONOlaw}) \emph{a forteriori}, because these ground state energies are $<0$ and they 
satisfy (\ref{vMONOlaw}). 
 Also the putatively optimal LJ$_N$ energy data at \cite{CCD} do satisfy (\ref{vMONOlaw}), and are $<0$, so 
they do satisfy (\ref{wMONOlaw}) as well. 

 For all the cases $Z\neq 0$ we need to verify that assumption (A) is satisfied to be able to use (\ref{wMONOlaw}) as 
a monotonicity test. 
 A sufficient condition for all $N-1$-monomer subclusters of $\cC^{(N)}_{\rm min}$ 
 to have non-positive average Lennard-Jones pair energy is that for all $i\neq j$ one has $|\rV_i-\rV_j|\geq 1$; 
note, however, that this is not a necessary condition unless $N=3$.
 This condition is easily checked for putatively optimal clusters found through computer experiments.

 We first inspected the pairwise distances in the putatively optimal LJ$_N$ clusters with $N\in\{3,...,150\}$ 
at \cite{CCD}, which revealed that they all are $\geq 1$, and this result implies that assumption (A) 
is satisfied for all putatively optimal LJ$_N$ clusters in this range of $N$ values.
 This observation implies furthermore that assumption (A) is satisfied for lists of local LJAT$_N$ minimizers generated in the
manner described above when $|Z|$ is sufficiently small.
 How small is ``sufficiently small''?
  Inspection of the pairwise distances in these LJAT$_N$ clusters with $N\in\{3,...,150\}$ revealed that
for the physically normal regime $Z\in\{\frac{n}{10}\}_{n=1,...,20}$ one has $|\rV_i-\rV_j|\geq 1$ when $i\neq j$.
 Interestingly, for the hypothetical regime of negative $Z$ we found that  \\
$\min_{i\neq j}|\rV_i-\rV_j|< 1$ when $Z=-1$, although $|\rV_i-\rV_j|\geq 1$ when $i\neq j$ if $Z=-0.1$.
 This result does not mean that (A) is not satisfied if $Z=-1$, but that further checking is required.
 In the following analysis we only consider lists with $Z>0$.

 Since the local LJAT$_N$ energy minimizers with our selection of $Z$ values in the interval $0<Z\leq 2$ satisfy (A), 
if indeed they are global optimizers they must display the monotonicity compatible with (\ref{wMONOlaw}). 
 We have verified that (\ref{wMONOlaw}) does hold (in the sense of discrete forward derivative) 
for the local LJAT$_N$ energy minimizers with $N\in\{3,...,149\}$
when $Z\in\{\frac{n}{10}\}_{n=1,...,10}$.

 However, for $Z\in \{\frac{n}{10}\}_{n=11,...,20}$
the energies of the locally energy-minimizing LJAT$_N$ clusters failed the monotonicity test (\ref{wMONOlaw})
for certain $N$-values, which means that the local LJAT$_N$ minimizers found in the neighborhood of the putatively optimal 
(i.e.~global) LJ$_N$ minimizers that violate (\ref{wMONOlaw}) when $Z\in\{\frac{n}{10}\}_{n=11,...,20}$ 
are not themselves globally minimizing LJAT$_N$ clusters.
 We suspect that for these $Z$ values also some of our locally energy-minimizing LJAT$_N$ clusters that
do satisfy the monotonicity test (\ref{wMONOlaw}) are not globally energy-minimizing.
 The true LJAT$_N$ optimizers for these $N$-values are not in the 
algorithm's basin of attraction that includes the pertinent LJ$_N$ optimizer.
 Thus a search via basin-hopping \cite{lis87,WalesDoye1997} is indicated, and we hope to report on the results in future work.

 Curiously, the sets of ``failing $N$-values'' do not seem to hint at any systematic pattern.
 For example, increasing $Z$ from 1 (no failures) to 1.5 in steps of 0.1, 
the locally energy-minimizing LJAT$_N$ clusters produced by our method
 violate the monotonicity law  (\ref{wMONOlaw}) at the following $N$-values:
$N\in\{136,144,147\}$ when $Z=1.1$, $N\in\{144,147\}$ when $Z=1.2$, $N = 135$ when $Z=1.3$, 
$N\in\{38,77,147\}$ when $Z=1.4$, $N\in\{38,41,67,84\}$ when $Z=1.5$. 
 When $Z$ is further increased in steps of $0.1$, the number of failures jumps up but then varies slowly at first:
$9$ failures for $Z=0.6$, $10$ for $Z=0.7$, $11$ for $Z=0.8$, $13$ for $Z=0.9$.
 By the time one reaches $Z=2$ there are already 21 $N$-values at which  (\ref{wMONOlaw}) fails;
we refrain from listing all those, yet we find it worthy to note that of the ``failing $N$-values''
in the five sets labeled with $Z\in\{1.1,...,1.5\}$ only  $N=135$ is in the set of failing $N$ values at $Z=2$.

 Since the trimeric Axilrod--Teller interaction competes with the dimeric Lennard-Jones interaction, the locally
energy-minimizing LJAT$_N$ clusters that we found in the neighborhood of the (putatively) optimal LJ$_N$ clusters cannot 
themselves also be LJ$_N$ local minima. 
 For ``sufficiently large'' $|Z|$ the LJAT$_N$ clusters should therefore fail the dimeric monotonicity law 
(\ref{vMONOlaw}); recall that for ``sufficiently small'' $|Z|$ the dimeric monotonicity law (\ref{vMONOlaw}) will be
satisfied, by continuity.
 And so, once again, how ``large'' is ``sufficiently large''? 
 Interestingly, we found that the first failure to obey (\ref{vMONOlaw}) happens for $Z=0.8$, when the discrete
forward derivative of the $N=74$ cluster is just barely negative. 
 When $Z=0.9$ there are already nine values of $N$ for which (\ref{vMONOlaw}) is violated, namely when 
$N\in\{68,74,81,85,97,113,122,130,134\}$. 
 For $Z\in \{\frac{n}{10}\}_{n=1,...,7}$ the average Lennard-Jones pair energy of the LJAT$_N$ clusters 
satisfies the monotonicity law (\ref{vMONOlaw}), although this law has been derived for the true LJ$_N$ optimizers. 
 Apparently their distortions from optimality are so small that none of the configurations fail the average 
pair-energy monotonicity test for optimal Lennard-Jones clusters.
 These findings may be seen as complementary to our results reported in \S\ref{sec:tightness}, where we inquired into
the number of non-global local minimizers that pass the monotonicity law (\ref{vMONOlaw}). 

 The fact that the first violations of (\ref{vMONOlaw}) by the dimeric Lennard-Jones contribution to the total
energy of the locally energy-minimizing LJAT$_N$ clusters occur already for $Z=0.8$, while the first violations
of (\ref{wMONOlaw}) by the total energy of these LJAT$_N$ clusters occur when $Z=1.1 $ 
suggests that one should not expect any noticeable correlations between the $N$ values of the pertinent failures.
 Indeed, the sets of $N$-values for which (\ref{vMONOlaw}) is violated by the Lennard-Jones contribution to the
LJAT$_N$ cluster energy do not foreshadow the sets of $N$-values at which the monotonicity (\ref{wMONOlaw}) is going 
to be violated. 

\section{Conclusions}
\label{sec:FIN}

Our monotonicity law (\ref{vMONOlaw}) applies to the optimal average $K$-body energy of $N$-body clusters 
that interact exclusively through permutation-invariant $K$-body potentials that admit a global minimizer for each $N\geq K$.
 When $K=2$ it reduces to the monotonicity law previously established in \cite{KieJCP}.
 Our monotonicity law (\ref{wMONOlaw}) generalizes (\ref{vMONOlaw}) to a family of cluster models with an additive mix of such 
$k$-body potentials, $1\leq k\leq K$ with $K\geq 2$ fixed, {that satisfy} the additional assumption: \\ 
\ \hspace*{0.5cm} (A): There is a fixed $\kappa\in\{2,...,K\}$ such that for all $N\geq K+1$, all nontrivial \\
\ \hspace*{0.75cm} \ $k$-body energies of each $N-1$-body subcluster of the $N$-body ground state \\
\ \hspace*{0.75cm} \ $\cC^{(N)}_{\rm min}$ are non-positive if $k<\kappa$ and non-negative if $k > \kappa$. \\
 We have listed several examples, some old and some new, of cluster models that must satisfy our monotonicity 
inequalities, some directly relevant to studies of molecular clusters, others inspired by them, and yet others inspired by 
the Thomson problem.
	 The main message of our paper is that our monotonicity inequalities furnish 
useful necessary conditions for optimality that can be used to test lists of putatively optimal cluster data.

 We have complemented our mathematical analysis with some empirical {studies}. 
 In one of these empirical studies we inquired into how sharp our test criterion (\ref{vMONOlaw}) is by counting the number of 
local minima for Lennard-Jones clusters that pass the monotonicity test (\ref{vMONOlaw}). 
 We found that as $N$ increases numerous local minima exist that pass the test, {yet} the test does
not seem easy to improve --- which we also showed. 
 This finding may be seen as another illustration for why these optimization problems fall in the category NP-hard.

{Incidentally, our tightness study reflects the existence of several competing sequences of local energy minimizers, 
with bifurcations among them, that reveal some interesting patterns in the energy landscape of low-lying energy configurations.
 This observation may help to guide further investigations of the competition between structural families.}

 In another empirical study we considered $N$-body clusters, with $N\in\{3,...,150\}$, whose monomers interact with an additive
mix of a dimeric Lennard-Jones and a trimeric Axilrod--Teller potential, with the AT amplitude $Z$ as parameter. 
 For $Z=0$ the problem reduces to the problem of finding the optimal Lennard-Jones clusters, which we confidently assume
have been found and listed at \cite{CCD}. 
 We then computed the local energy minimizers of the LJAT problem in the vicinity of the $Z=0$ optimizers
as a function of $Z$, using steps of $\triangle Z = 0.1$.
 By continuity arguments one may expect that for small $|Z|$ this procedure will find the truly energy-optimizing LJAT$_N$ clusters,
which therefore should satisfy the monotonicity law (\ref{wMONOlaw}). 
 We found that  (\ref{wMONOlaw})  is obeyed for when $0< Z \leq 1.0$, but for $1.1\leq Z\leq 2$ we found 
that some of the locally energy-minimizing LJAT$_N$ clusters violated (\ref{wMONOlaw}), demonstrating that the true global minima
are not in the basin of attraction that contains the $Z=0$ optimizers. 
 We verified numerically that assumption (A) is satisfied to guarantee that our inequality (\ref{wMONOlaw}) furnishes a 
valid test.

 We just mentioned that we empirically verified assumption (A) for all our LJAT$_N$ data. 
 This analysis sufficed for us to vindicate the test of our LJAT$_N$ data against the monotonicity law (\ref{wMONOlaw}). 
 However, no check of empirical lists of putatively optimal configurations proves for sure that the true 
LJAT$_N$ optimizers satisfy assumption (A), and this for \emph{all} $N\geq 3$. 
 A rigorous proof that all $|\rV_i-\rV_j|\geq 1$ in the optimal LJAT$_N$ clusters for all $N\geq 3$ would take care of 
this open issue, yet we don't expect to see a rigorous proof anytime soon.

 {It is not even rigorously known whether}
$\min_{i\neq j} |\rV_i-\rV_j|\geq 1$ in the optimal LJ$_N$ clusters for all $N\geq 2$.
 Yet there is strong empirical and theoretical evidence that this is a lower bound on the minimal distance in LJ$_N$,
independently of $N$; in fact, we expect that ``$\geq$'' can be replaced with ``$>$.''

{First of all, with the help of the virial theorem it can be shown} that for all $N\geq 2$ one
has $r_{\mbox{\tiny{min}}}^{}(N) \leq 2^\frac16$, the minimizing separation for the dimer potential in our units.
 Since $N=2$ is included, this upper bound is sharp; in fact, it is easily seen that
\begin{equation}
 r_{\mbox{\tiny{min}}}^{}(N) = 2^\frac16 \approx 1.122462048; \quad N\in\{2,3,4\}.
\end{equation}

 Second, there is also some theoretical evidence from studies of crystal structures in the thermodynamic limit $N\to\infty$.
 In \cite{JonesIngham} the minimal pairwise distance was computed for spatially unbounded simple cubic,
bcc and fcc Lennard-Jones crystals, by minimizing their energy per monomer.
 They found that the fcc crystal has the lowest energy per particle among these three regular lattices,
and for a long time it was thought that fcc is  the optimal {crystal structure} in the limit $N\to\infty$.
 Now we know that fcc is the optimal \emph{standard lattice} structure, with a minimal pairwise distance 
that can then be computed from the results in \cite{JonesIngham} as 
\begin{equation}
 r_{\mbox{\tiny{min}}}^{\mbox{\tiny{fcc}}} \approx 1.090172.
\end{equation}
 However, recently it was found empirically  \cite{ZschornakETal}, 
and then proved rigorously \cite{BST}, that the hcp crystal structure is the true
crystalline ground state configuration in the thermodynamics limit.
 Note {that the hcp crystal} is not a regular lattice in the same sense as fcc, bcc, and simple cubic crystals.
 Using the results of \cite{BCPS}, the minimal pairwise distance in the optimal crystal structure 
can be computed as
\begin{equation}
 r_{\mbox{\tiny{min}}}^{\mbox{\tiny{hcp}}}
 \approx 1.090167,
\end{equation}
which is slightly smaller than $ r_{\mbox{\tiny{min}}}^{\mbox{\tiny{fcc}}}$.

 Third, we inspected the putatively optimal Lennard-Jones clusters LJ$_N$ at \cite{CCD} for $N\leq 150$ and 
found that the smallest minimal distance  $r_{\mbox{\tiny{min}}}^{}(N)$ occurs for $N=71$, 
\begin{equation}
 r_{\mbox{\tiny{min}}}^{}(71) \approx 1.028758632.
\end{equation}

 So far there are no decisive theoretical results for the vast intermediate regime between
these values at the extreme ends of the $N$ scale.
 In \cite{Xue} it has been shown that there is an $N$-independent lower bound 
on the smallest pairwise distance in any globally energy-minimizing Lennard-Jones cluster, 
and the following lower bound was given (rescaled into our units):
 $r_{\mbox{\tiny{min}}}^{}(N) \geq 1/2^\frac{5}{6}\approx 0.561231$.
 Subsequent papers by \cite{Vinko}, then \cite{SchachingerETal}, 
and more recently \cite{Yuhjtman} have improved the theoretical lower bound on this quantity to currently
 $r_{\mbox{\tiny{min}}}^{}(N) \geq 0.767764$.
 Clearly, this best known bound is still far away from the suspected bound
$r_{\mbox{\tiny{min}}}^{}(N) \geq 1$. 
\medskip

{A semi-final remark: Our empirical studies are meant not only to illustrate the usefulness of the test criteria
obtained in this paper;
they are also meant to inspire readers to apply them to their own lists of data, and to try to find alternative or better criteria
for optimality. 
 In particular, as noted by the anonymous referee, it is reasonable to ask whether the $N$-dependence of the variance over all 
the $K$-body energies in an $N$-body cluster with pure $K$-body interactions furnishes a useful test criterion for optimality.
 We don't know the answer, but this question certainly merits further analysis.}

 Our final remark highlights the fact that the monotonicity of the ground state energy is 
useful for other purposes.
 About 100 years ago, and thus long before powerful computers became available that could produce lists of putatively 
optimal clusters, special cases of the $K=2$ version of (\ref{vMONOlaw}) have been proved \cite{Fekete}, \cite{PS},
in a very different context; see Appendix B for some of the history of the $K=2$ version of (\ref{vMONOlaw}).
 In this vein we expect that the monotonicity laws derived in this paper will be useful also in contexts other than testing
 computer-generated lists of putatively optimal energies.

\bigskip
\noindent
\textbf{ACKNOWLEDGEMENT}: We thank Bhargav Narayanan for clarifying conversations about hypergraphs, and
Neil Sloane for pointing out \cite{TeoSloane}.
 We also thank the anonymous referee for the interesting comments.

\newpage
\begin{appendices}
\section{Hypergraph-theoretical input}\label{sec:HyperGid}
\vspace{-.2truecm}

\subsection{Hypergraph terminology}

{A \emph{non-directed hypergraph} is a pair $(\cV,\cE)$ where $\cV$ is an unordered set of points 
called \emph{vertices} and $\cE$ is a set of unordered subsets of $\cV$, called \emph{hyperedges}. 
 Any unordered subset of $k\in \{1,...,N\}$ vertices of such a hypergraph with $N$ vertices 
can be ``joined into a hyperedge,'' here called a $k$-hyperedge, for short.
 Special names are available for the 1-hyperedge (just a vertex) and the 2-hyperedge (the usual edge of a non-empty graph). 
 The number $k$ of vertices in a $k$-hyperedge suggests to give $\cE$ an ordering, or a grading, but we won't need it.}

 In the following we are only interested in non-directed hypergraphs, and so the adjective ``non-directed'' will henceforth
be dropped. 
 Also, if we speak of ``subsets of vertices,'' we tacitly mean ``unordered subsets.''

 Typically not all possible subsets of $\cV$ occur in $\cE$, but if they do (i.e. if $\cE$ is the powerset of $\cV$), 
the hypergraph is called \emph{complete}.
 For each $k\leq N$ there are ${\genfrac{(}{)}{0pt}{1}{N}{k}}$ different $k$-hyperedges in such a hypergraph.
 If $\cE$ consists entirely of $k$-hyperedges with a fixed $k=K$, the hypergraph is called
\emph{$k$-uniform}, and such a $k$-uniform hypergraph is \emph{complete} if $\cE$ contains all subsets with $k=K$ 
vertices.

 A hypergraph is \emph{weighted} if all hyperedges are assigned a weight.
 Usually a weight is non-negative, but we allow real weights (perhaps a better terminology would be ``charged'').
 Including vanishing weights allows one to work with complete hypergraphs by default.

\subsection{Clusters as hypergraphs}

 In \cite{KieJCP} we identified clusters whose monomers interact solely pairwise with complete weighted graphs. 
 More generally, any cluster of $N\geq K\geq 2$ monomers that properly interact through permutation-invariant 
irreducible $k$-body potentials $V_k$ for some (or all) $k$ satisfying $2\leq k\leq K$, 
and whose individual monomer states in addition may be assigned nontrivial one-body potentials $V_1$, 
can be identified with a complete weighted hypergraph --- provided one stipulates that for any 
$k\in\{1,...,N\}$ for which no nontrivial $V_k$ appears in the model, in particular for all $k\in\{K+1,...,N\}$, 
the trivial $k$-body interaction $V_k\equiv 0$ is assigned as the weight to all those $k$-hyperedges. 

 More to the point, the states of the $N$ monomers in a cluster are identified with the $N$ vertices of a hypergraph. 
 For our cluster models the permutation-invariant, irreducible $k$-body interactions $V_k$, and the single-monomer
potential $V_1$, if nonvanishing, define the relevant weighted $k$-hyperedges of the hypergraph together with their
nontrivial weights $V_k$, while all other $k$-hyperedges are assigned trivial weights and are, therefore, irrelevant
--- they are included merely for later notational convenience.
 This setup includes all cluster models covered in the main part of this work, whether the monomers interact
through an additive mixture of permutation-symmetric many-body potentials $V_k$ for various different $k\leq K$,
and are additionally equipped with a nontrivial one-monomer potential $V_1$, or whether only $k$-body interactions 
for a single value $k=K\geq 2$ feature. 
 In the latter case all $V_{k\neq K}\equiv 0$, and the complete weighted hypergraph is equivalent to 
a $K$-uniform complete weighted hypergraph. 

 With these stipulations we may simply replace $K$ by $N$ in (\ref{WofMULTIkAVEsum}).
 The total energy (\ref{WofMULTIkAVEsum}) of an $N$-monomer cluster
is thus seen to be the sum of all the hyperedge weights (trivial and nontrivial) over all the hyperedges of 
a weighted complete hypergraph.

\subsection{A family of hypergraph-theoretical identities}

 Let $N>K\geq 2$. 
 Consider the contribution of the $k$-body potential with $k\in\{1,...,K\}$ to $W$ given in  
(\ref{WofMULTIkAVEsum}).
 Without loss of generality we may assume that $V_k\not\equiv 0$ (when $V_k\equiv 0$ the following is trivially true). 
 Define the $\ell$-th sub-hypergraph obtained from the complete hypergraph with $N$ vertices 
by removing the $\ell$-th vertex and, hence, with it all $k$-hyperedges containing the $\ell$-th vertex.
 If we now sum the hyperedge-weighting function over all $k$-hyperedges of that $\ell$-th sub-hypergraph, 
and then sum the generally $\ell$-dependent result of this summation over $\ell$, i.e.~over all vertices 
of the hypergraph with $N$-vertices, then all $k$-hyperedges appear in the final result $(N-k)$ times.
 Thus we overcount the $k$-th contribution to the total energy $W({\cal C}^{(N)})$ by a factor $(N-k)$, 
i.e. we get the family of hypergraph-theoretical identities
\begin{equation}\label{eq:WidANYk}
\sum_{1\leq \ell\leq N}\
\sum\; \cdots\!\!\sum_{\hskip-1.3truecm \begin{subarray}{c} 1\leq i_1<\cdots<i_k \leq N 
\\ \!\! \ell\not\in \{i_1,...,i_k\}       \end{subarray}                      } 
V_k(S_{i_1},...,S_{i_k})
=
(N-k)
\sum\; \cdots\!\!\sum_{\hskip-1.3truecm 1\leq i_1<\cdots<i_k \leq N} V_k(S_{i_1},...,S_{i_k})
\end{equation}
valid for all $k\in\{1,...,K\}$.

 Identity (\ref{eq:WidANYk}) can be written more elegantly as follows.
 Dividing both sides of (\ref{eq:WidANYk}) by ${\genfrac{(}{)}{0pt}{1}{N}{k}}(N-k)$, viz.
by $N {\genfrac{(}{)}{0pt}{1}{N-1}{k}}$, and recalling the definition (\ref{eq:AveVk})
of the average $k$-body energy of an $N$-monomer configuration,  (\ref{eq:WidANYk}) becomes
\begin{equation}\label{eq:AveVkID}
\forall k\in\{1,...,K\}:\ 
\langle V_k\rangle \big(\cC^{(N)}\big) 
= \frac{1}{N}\sum_{1\leq\ell\leq N} 
 \langle V_k\rangle \big(\cC^{(N)}\backslash\{S_\ell\}\big).
\end{equation}

 We note that the special case $k=K=2$ of (\ref{eq:AveVkID}) is identity (9) in \cite{KieJCP}.

\section{On the history of the monotonicity of the optimal average pair energy}
 
 The first publication to advocate the use of the monotonicity of the optimal average pair energy of $N$-body
clusters whose monomers interact solely with a certain class of permutation-symmetric pair potentials,
 as a necessary condition for optimality to test putative global minima, 
seems to be \cite{KieJSP}. 
 The most general class so far of $N$-body systems with pair interactions for which the monotonicity has 
been established is the subsequent article \cite{KieJCP}.
 These monotonicity results are included as special case of our $K$-body result (\ref{vMONOlaw}), which in
turn is included in (\ref{wMONOlaw}).

 We find it worthy of note that the monotonic increase of the average pair energy of $N$-body ground states was first proved
in a completely different context, though, and this about 100 years ago!
 For the special case of logarithmic pair interactions between points in a compact infinite subset $D\subset\Cset$, Fekete 
(see \cite{Fekete}, $\S 1$, Thm. I$^*$) established the corresponding $K=2$ special case of (\ref{vMONOlaw}).  
 He didn't express his results in this terminology, though.
 Fekete was interested in the distribution of the zeros of complex polynomials with integer coefficients, and
used his monotonicity theorem to establish the existence of the so-called ``transfinite diameter of $D$,'' a limit of the
\emph{geometric mean distance} between the zeros of the order-$N$ polynomials when $N\to\infty$. 
 Taking a logarithm of Fekete's formulas converts this narrative into one of clusters with logarithmic pair interactions.

 Fekete's result was subsequently generalized by Polya and Szeg\H{o} \cite{PS} to points that interact pairwise with the
fundamental solution of the Laplacian in $\Rset^3$. 
 Their monotonicity results have in turn been generalized to the family of Riesz kernels of potential theory in $\Rset^d$,
$d>2$, for the parameter interval $s\in(0,d)$; see  Landkof's book \cite{Landkof}, Ch.II, \S 3, No.12, p.160; 
and see \S 4, No.15, p. 169 for Fekete's result.

 The monotonicity of the optimal average pair energy of clusters with a larger class of pair interactions, yet still 
expressible as functions of the particle positions only, was subsequently rediscovered in \cite{KieRMP}
(see App.A, Proof of Prop.7, p.1189). 
 More precisely, the proof is stated for 
point particle systems with symmetric and lower semi-continuous pair interactions in compact subsets $D\subset\Rset^d$.
 
 In all the works \cite{Fekete}, \cite{PS}, \cite{KieRMP},
the monotonicity of the optimal average pair energy, together with a uniform bound on the average pair energy, was employed
to establish its limit as $N\to\infty$.  

 By analogy, our monotonicity results (\ref{vMONOlaw}), in concert with corresponding uniform bounds on the average $K$-body
energy, will establish the limit $N\to\infty$ of the pertinent average $K$-body energy. 

\end{appendices}

\smallskip
\hrule
\smallskip

{\bf Conflict-of-Interest Statement}: The authors have no conflict of interest.

\newpage
  
\bibliographystyle{plain}
\setstretch{0.87}
\bibliography{CLUSTER}

\begin{thebibliography}{10}

\bibitem{Adip}
A.~B. Adip.
\newblock {NP}-hardness of the cluster minimization problem revisited.
\newblock {\em J. Phys. A: Math. Gen.}, \textbf{38}:8487--8492, (2005).

\bibitem{AltschulerETal}
E.L. Altschuler, T.J. Williams, E.R. Ratner, R.~Tipton, R.~Stong, F.~Dowla, and
  F.~Wooten.
\newblock Possible global minimum lattice configurations for {T}homson’s
  problem of charges on the sphere.
\newblock {\em Phys. Rev. Lett.}, \textbf{78}:2681--2685, (1997).

\bibitem{AsenjoSWF13}
D.~Asenjo, J.~D. Stevenson, D.~J. Wales, and D.~Frenkel.
\newblock Visualizing basins of attraction for different minimization
  algorithms.
\newblock {\em J. Phys. Chem. B}, \textbf{117}:12717--12723, (2013).

\bibitem{AxilrodTeller}
B.~M. Axilrod and E.~Teller.
\newblock Interaction of the van der {W}aals type between three atoms.
\newblock {\em J. Chem. Phys.}, \textbf{11}:299--300, (1943).

\bibitem{BRb}
C.~Barr\'on-Romero.
\newblock Optimal clusters.
\newblock \url{https://academicos.azc.uam.mx/cbr/OptClusters/comMRLJMO_01.htm}.

\bibitem{BRa}
C.~Barr\'on-Romero.
\newblock The olj13$\underline{\ }$n13ic cluster is the global minimum cluster
  of {L}ennard-{J}ones potential for 13 particles.
\newblock In {\em 2022 IEEE 3rd International Conference on Electronics,
  Control, Optimization and Computer Science (ICECOCS), Dec.01-02, 2022}. IEEE,
  (2022).

\bibitem{Beck}
J.~Beck.
\newblock Sums of distances between points on a sphere --- an application of
  the theory of irregularities of distribution to discrete geometry.
\newblock {\em Mathematica}, \textbf{31}:33--41, (1984).

\bibitem{BST}
L.~B\'etermin, L.~\v{S}amaj, and Trav\v{e}nec.
\newblock Three-dimensional lattice ground states for {R}iesz and
  {L}ennard-{J}ones--type energies.
\newblock {\em Stud. Appl. Math.}, \textbf{150}:69--91, (2023).

\bibitem{BezdekKhan}
K.~Bezdek and M.~A. Khan.
\newblock Contact numbers for sphere packings.
\newblock In G.~Ambrus, I.~B\'ar\'any, J.~K. B\"or\"oczky, G.~Fejes~T\'oth, and
  J.~Pach, editors, {\em New Trends in Intuitive Geometry}, volume \textbf{27}
  of {\em Bolyai {S}ociety {M}athematical {S}tudies}, pages 25--48. Springer
  Verlag, New York, (2018).

\bibitem{Bjoerck}
G.~Bj{\"o}rck.
\newblock Distributions of positive mass, which maximize a certain generalized
  energy integral.
\newblock {\em Ark. Mat.}, \textbf{3}:255--269, (1956).

\bibitem{BCNTa}
M.~Bowick, A.~Cacciuto, D.~R. Nelson, and A.~Travesset.
\newblock Crystalline order on a sphere and the generalized {T}homson problem.
\newblock {\em Phys. Rev. Lett.}, \textbf{89}:185502,1--4, (2002).

\bibitem{BCNTb}
M.~Bowick, A.~Cacciuto, D.~R. Nelson, and A.~Travesset.
\newblock Crystalline particle packings on a sphere with long-range power-law
  potentials.
\newblock {\em Phys. Rev. B}, \textbf{73}:024115,1--16, (2006).

\bibitem{Broyden70}
C.~G. Broyden.
\newblock The convergence of a class of double-rank minimization algorithms 1.
  {G}eneral considerations.
\newblock {\em J. Inst. Math. Appl.}, \textbf{6}:76--90, (1970).

\bibitem{BCPS}
A.~Burrows, S.~Cooper, E.~Pahl, and P.~Schwerdtfeger.
\newblock Analytical methods for fast converging lattice sums for cubic and
  hexagonal close-packed structures.
\newblock {\em J. Math. Phys.}, \textbf{61}:123503, 1--35, (2020).

\bibitem{CalvoDW01b}
F.~Calvo, J.~P.~K. Doye, and D.~J. Wales.
\newblock Quantum partition functions from classical distributions: Application
  to rare-gas clusters.
\newblock {\em J. Chem. Phys.}, 114:7312--7329, 2001.

\bibitem{Crocker10}
J.~C. Crocker.
\newblock Turning away from high symmetry.
\newblock {\em Science}, 327:535--536, (2010).

\bibitem{Doye}
J.~P.~K. Doye.
\newblock {L}ennard-{J}ones clusters.
\newblock \url{http://doye.chem.ox.ac.uk/jon/structures/LJ.html}.

\bibitem{DoyeC02}
J.~P.~K. Doye and F.~Calvo.
\newblock Entropic effects on the structure of lennard-jones clusters.
\newblock {\em J. Chem. Phys.}, 116:8307--8317, 2002.

\bibitem{DoyeMillerWalesA}
J.~P.~K. Doye, M.~A. Miller, and D.~J. Wales.
\newblock The double-funnel energy landscape of the 38-atom {L}ennard-{J}ones
  cluster.
\newblock {\em J. Chem. Phys.}, \textbf{110}:6896--6906, (1999).

\bibitem{DoyeMillerWalesB}
J.~P.~K. Doye, M.~A. Miller, and D.~J. Wales.
\newblock Evolution of the potential energy surface with size for
  {L}ennard-{J}ones clusters.
\newblock {\em J. Chem. Phys.}, \textbf{111}:8417--8428, (1999).

\bibitem{doyew92}
J.~P.~K. Doye and D.~J. Wales.
\newblock Systematic investigation of the structures and rearrangements of
  6-atom clusters bound by a model anisotropic potential.
\newblock {\em J. Chem. Soc. Faraday Trans.}, \textbf{88}:3295--3304, (1992).

\bibitem{doyew96a}
J.~P.~K. Doye and D.~J. Wales.
\newblock The effect of the range of the potential on the structure and
  stability of simple liquids - from clusters to bulk, from sodium to c-60.
\newblock {\em J. Phys. B}, 29:4859--4894, 1996.

\bibitem{DoyeWales2002}
J.~P.~K. Doye and D.~J. Wales.
\newblock Saddle points and dynamics of {L}ennard-{J}ones clusters, solids, and
  supercooled liquids.
\newblock {\em J. Chem. Phys.}, \textbf{116}:3777--3788, (2002).

\bibitem{Fekete}
M.~Fekete.
\newblock {\"U}ber die {V}erteilung der {W}urzeln bei gewissen algebraischen
  {G}lei\-chungen mit ganzzahligen {K}oeffizienten.
\newblock {\em Math. Z.}, \textbf{17}:228--249, (1923).

\bibitem{Fletcher70}
R.~Fletcher.
\newblock A new approach to variable metric algorithms.
\newblock {\em Comput. J.}, \textbf{13}:317--322, (1970).

\bibitem{FormanCameron}
Y.~Forman and M.~K. Cameron.
\newblock Modeling aggregation processes of {L}ennard-{J}ones particles via
  stochastic networks.
\newblock {\em J. Statist. Phys.}, \textbf{168}:408--433, (2017).

\bibitem{GlotzerE11}
S.~C. Glotzer and M.~Engel.
\newblock Complex order in soft matter.
\newblock {\em Nature}, \textbf{471}:309--310, (2011).

\bibitem{GlotzerS07}
S.~C. Glotzer and M.~J. Solomon.
\newblock Anisotropy of building blocks and their assembly into complex
  structures.
\newblock {\em Nature Mater.}, \textbf{6}:557--562, (2007).

\bibitem{Goldberg}
M.~Goldberg.
\newblock Maximizing the smallest triangle made by {$N$} points in a square.
\newblock {\em Math. Mag.}, \textbf{45}:135--144, (1972).

\bibitem{Goldfarb70}
D.~Goldfarb.
\newblock A family of variable-metric methods derived by variational means.
\newblock {\em Math. Comput.}, \textbf{24}:23--26, (1970).

\bibitem{GC}
T.~Gregor and R.~Car.
\newblock Minimization of the potential energy surface of {L}ennard-{J}ones
  clusters by quantum optimization.
\newblock {\em Chem. Phys. Lett.}, \textbf{412}:125--130, (2005).

\bibitem{Harborth}
H.~Harborth.
\newblock L\"osung zu {P}roblem 664{A}.
\newblock {\em Elem. Math.}, \textbf{29}:14--15, (1974).

\bibitem{holmes17}
M.~Holmes-Cerfon.
\newblock Sticky-sphere clusters.
\newblock {\em Annual Rev. Cond. Matter Phys.}, \textbf{8}:77--98, (2017).

\bibitem{holmes13}
M.~Holmes-Cerfon, S.~J. Gortler, and M.~P. Brenner.
\newblock A geometrical approach to computing free-energy landscapes from
  short-ranged potentials.
\newblock {\em Proc. Nat. Acad. Sci.}, \textbf{110}:E5--E10, (2013).

\bibitem{holmes16}
M.~C. Holmes-Cerfon.
\newblock Enumerating rigid sphere packings.
\newblock {\em SIAM Rev.}, \textbf{58}:229--244, (2016).

\bibitem{hoy15}
R.~S. Hoy.
\newblock Structure and dynamics of model colloidal clusters with short-range
  attractions.
\newblock {\em Phys. Rev. E}, \textbf{91}:012303, 1--7, (2015).

\bibitem{hoy_structure_2012}
R.~S. Hoy, J.~Harwayne-Gidansky, and C.~S. O'Hern.
\newblock Structure of finite sphere packings via exact enumeration:
  {{Implications}} for colloidal crystal nucleation.
\newblock {\em Phys. Rev. E}, \textbf{85}:051403,1--15, (2012).

\bibitem{hoy_minimal_2010}
R.~S. Hoy and C.~S. O'Hern.
\newblock Minimal energy packings and collapse of sticky tangent hard-sphere
  polymers.
\newblock {\em Phys. Rev. Lett.}, \textbf{105}:068001,1--4, (2010).

\bibitem{Jellinek99}
J.~Jellinek, editor.
\newblock {\em Theory of atomic and molecular clusters}.
\newblock Springer-Verlag, Heidelberg, 1999.

\bibitem{RoyBook}
R.~L. Johnston.
\newblock {\em Atomic and molecular clusters}.
\newblock Taylor and Francis, London and New York, (2002).

\bibitem{JonesIngham}
J.~E. Jones and A.~E. Ingham.
\newblock On the calculation of certain crystal potential constants, and on the
  cubic crystal of least potential energy.
\newblock {\em Proc. Royal Soc. London, Ser.A}, \textbf{107}:636--653, (1925).

\bibitem{kallus17}
Y.~Kallus and M.~Holmes-Cerfon.
\newblock Free energy of singular sticky-sphere clusters.
\newblock {\em Phys. Rev. E}, \textbf{95}:022130, 1--18, (2017).

\bibitem{KieJSP}
M.~K.-H. Kiessling.
\newblock A note on classical ground state energies.
\newblock {\em J. Statist. Phys.}, \textbf{136}:275--284, (2009).

\bibitem{KieRMP}
M.~K.-H. Kiessling.
\newblock The {V}lasov continuum limit for the classical microcanonical
  ensemble.
\newblock {\em Rev. Math. Phys.}, \textbf{21}:1145--1195, (2009).

\bibitem{KieJCP}
M.~K.-H. Kiessling.
\newblock Testing {L}ennard-{J}ones clusters for optimality.
\newblock {\em J. Chem. Phys.}, \textbf{159}:014301, 1--6, (2023).

\bibitem{KieYi}
M.~K.-H. Kiessling and R.~Yi.
\newblock Hilbert's `{M}onkey {S}addle' and other curiosities in the
  equilibrium problem of three point particles on a circle for repulsive power
  law forces.
\newblock {\em J. Dyn. Diff. Eq.}, \textbf{32}:233--256, (2020).

\bibitem{KPS}
J.~Koml\'os, J.~Pintz, and E.~Szemer\'edi.
\newblock On {H}eilbronn's triangle problem.
\newblock {\em J. London Math. Soc.}, \textbf{24}:385--396, (1981).

\bibitem{LXH}
X.J. Lai, R.C. Xu, and W.Q. Huang.
\newblock Prediction of the lowest energy configuration for {L}ennard-{J}ones
  clusters.
\newblock {\em Sci. China Chem.}, \textbf{54}:985--991, (2011).

\bibitem{Landkof}
N.~S. Landkof.
\newblock {\em Foundations of Modern Potential Theory}, volume \textbf{180} of
  {\em Grundlehren der mathematischen Wissenschaften}.
\newblock Springer Verlag, Berlin, (1972).

\bibitem{Leary}
R.~H. Leary.
\newblock Global optima of {L}ennard-{J}ones clusters.
\newblock {\em J. Global Optim.}, \textbf{11}:35--53, (1997).

\bibitem{Lefmann}
H.~Lefmann.
\newblock On {H}eilbronn’s problem in higher dimension.
\newblock {\em Combinatorica}, \textbf{23}:669--680, (2003).

\bibitem{lis87}
Z.~Li and H.~A. Scheraga.
\newblock Monte {C}arlo-minimization approach to the multiple-minima problem in
  protein folding.
\newblock {\em Proc. Natl Acad. Sci. USA}, \textbf{84}:6611--6615, (1987).

\bibitem{lbfgs}
D.~Liu and J.~Nocedal.
\newblock On the limited memory bfgs method for large scale optimization.
\newblock {\em Math. Prog.}, \textbf{45}:503--528, (1989).

\bibitem{LocatelliSchoenA}
M.~Locatelli and F.~Schoen.
\newblock Efficient algorithms for large scale global optimization:
  {L}ennard-{J}ones clusters.
\newblock {\em Comput. Optim. Appl.}, \textbf{26}:173--190, (2003).

\bibitem{mackay62}
A.~L. Mackay.
\newblock A dense non-crystallographic packing of equal spheres.
\newblock {\em Acta Cryst.}, \textbf{15}:916--918, (1962).

\bibitem{MahoneyJ00}
M.~W. Mahoney and W.~L. Jorgensen.
\newblock A five-site model for liquid water and the reproduction of the
  density anomaly by rigid, nonpolarizable potential functions.
\newblock {\em J. Chem. Phys.}, \textbf{112}:8910--8922, (2000).

\bibitem{MengABM10}
G.~Meng, N.~Arkus, M.~P. Brenner, and V.~N. Manoharan.
\newblock The free-energy landscape of clusters of attractive hard spheres.
\newblock {\em Science}, \textbf{327}:560--563, (2010).

\bibitem{MoessnerRamirez}
R.~Moessner and A.~Ramirez.
\newblock Geometric frustration.
\newblock {\em Phys. Today}, \textbf{59}:24--29, (2006).

\bibitem{Morse29}
P.~M. Morse.
\newblock Diatomic molecules according to the wave mechanics. {I}{I}.
  {V}ibrational levels.
\newblock {\em Phys. Rev.}, \textbf{34}:57--64, (1929).

\bibitem{MuellerSbalzarini}
C.~L. M\"uller and I.~F. Sbalzarini.
\newblock Energy landscapes of atomic clusters as black box optimization
  benchmarks.
\newblock {\em Evol. Comput.}, \textbf{20}:543--573, (2012).

\bibitem{NBK}
R.~Nerattini, J.~S. Brauchart, and M.~K.-H. Kiessling.
\newblock Optimal {$N$}-point configurations on the sphere: ``{M}agic'' numbers
  and {S}male's 7th problem.
\newblock {\em J. Statist. Phys.}, \textbf{157}:1138--1206, (2014).

\bibitem{Nocedal80}
J.~Nocedal.
\newblock Updating quasi-{N}ewton matrices with limited storage.
\newblock {\em Mathematics of Computation}, \textbf{35}:773--782, (1980).

\bibitem{Northby}
J.~A. Northby.
\newblock Structure and binding of {L}ennard-{J}ones clusters: $13\leq n\leq
  147$.
\newblock {\em J. Chem. Phys.}, \textbf{87}:6166--6177, (1987).

\bibitem{PS}
G.~P\'olya and G.~Szeg\H{o}.
\newblock {\"U}ber den transfiniten {D}urchmesser ({K}apa\-zit\"ats\-konstante)
  von ebenen und r\"aumlichen {P}unktmengen.
\newblock {\em J. Reine Angew. Math.}, \textbf{165}:4--49, (1931).

\bibitem{Roth}
K.~F. Roth.
\newblock On a problem of {H}eilbronn.
\newblock {\em J. London Math. Soc.}, \textbf{26}:198--204, (1951).

\bibitem{SacannaICP10}
S.~Sacanna, W.~T.~M. Irvine, P.~M. Chaikin, and D.~J. Pine.
\newblock Lock and key colloids.
\newblock {\em Nature}, \textbf{464}:575--578, (2010).

\bibitem{SchachingerETal}
W.~Schachinger, B.~Addis, I.~M. Bomze, and F.~Schoen.
\newblock New results for molecular formation under pairwise potential
  minimization.
\newblock {\em Comput. Optim. Appl.}, \textbf{38}:329--349, (2007).

\bibitem{SchVdW}
K.~Sch{\"u}tte and B.~L. Van~der Waerden.
\newblock Das {P}roblem der dreizehn {K}ugeln.
\newblock {\em Math. Ann.}, \textbf{125}:325–334, (1953).

\bibitem{Shanno70}
D.~F. Shanno.
\newblock Conditioning of quasi-{N}ewton methods for function minimization.
\newblock {\em Math. Comput.}, \textbf{24}:647--656, (1970).

\bibitem{Shao}
X.~Shao.
\newblock The structures of the optimized {L}ennard-{J}ones clusters.
\newblock \url{https://chinfo.nankai.edu.cn/labintro_e.html}.

\bibitem{Smale}
S.~Smale.
\newblock Mathematical problems for the next century.
\newblock {\em Math. Intelligencer}, \textbf{20}:7--15, (1998).

\bibitem{StillWebA}
F.~H. Stillinger and T.~A. Weber.
\newblock Hidden structure in liquids.
\newblock {\em Phys. Rev. A}, \textbf{25}:978--989, (1982).

\bibitem{StillWebB}
F.~H. Stillinger and T.~A. Weber.
\newblock Packing structures and transitions in liquids and solids.
\newblock {\em Science}, \textbf{225}:983--989, (1984).

\bibitem{Sugano91}
S.~Sugano.
\newblock {\em Microcluster physics}.
\newblock Springer-Verlag, Berlin, 1991.

\bibitem{TeoSloane}
B.~K. Teo and N.~J.~A. Sloane.
\newblock Magic numbers in polygonal and polyhedral clusters.
\newblock {\em Inorg. Chem.}, \textbf{24}:4545--4558, (1985).

\bibitem{Thomson04}
J.~J. Thomson.
\newblock On the structure of the atom: an investigation of the stability and
  periods of oscillation of a number of corpuscles arranged at equal intervals
  around the circumference of a circle; with application of the results to the
  theory of atomic structure.
\newblock {\em Phil. Mag., Ser. 6}, \textbf{7}:237--265, (1904).

\bibitem{PhysRevE.97.043309}
L.~Trombach, R.~S. Hoy, D.~J. Wales, and P.~Schwerdtfeger.
\newblock From sticky-hard-sphere to {L}ennard-{J}ones-type clusters.
\newblock {\em Phys. Rev. E}, \textbf{97}:043309,1--10, (2018).

\bibitem{Vinko}
T.~Vink\'o.
\newblock Minimal inter-particle distance in atom clusters.
\newblock {\em Acta Cybern.}, \textbf{17}:105--119, (2005).

\bibitem{OPTIM}
D.~J. Wales.
\newblock {\tt OPTIM}: A program for geometry optimisation and pathway
  calculations.
\newblock http://www-wales.ch.cam.ac.uk/software.html.

\bibitem{wales90a}
D.~J. Wales.
\newblock Structural and topological consequences of anisotropic interactions
  in clusters.
\newblock {\em J. Chem. Soc. Faraday Trans.}, \textbf{86}:3505--3517, (1990).

\bibitem{wales93d}
D.~J. Wales.
\newblock Locating stationary-points for clusters in cartesian coordinates.
\newblock {\em J. Chem. Soc. Faraday Trans.}, \textbf{89}:1305--1313, (1993).

\bibitem{WalesBOOK}
D.~J. Wales.
\newblock {\em Energy Landscapes: With Applications to Clusters, Biomolecules
  and Glasses}.
\newblock Cambridge Univ. Press, Cambdridge, UK, (2004).

\bibitem{Wales10b}
D.~J. Wales.
\newblock Highlights: Energy landscapes of clusters bound by short-ranged
  potentials.
\newblock {\em ChemPhysChem}, \textbf{11}:2491--2494, (2010).

\bibitem{Wales13}
D.~J. Wales.
\newblock Surveying a complex potential energy landscape: Overcoming broken
  ergodicity using basin-sampling.
\newblock {\em Chem. Phys. Lett.}, \textbf{584}:1--9, (2013).

\bibitem{WalesDoye1997}
D.~J. Wales and J.~P.~K. Doye.
\newblock Global optimization by basin-hopping and the lowest energy structures
  of {L}ennard-{J}ones clusters containing up to 110 atoms.
\newblock {\em J. Phys. Chem. A}, \textbf{101}:5111--5116, (1997).

\bibitem{WalesDoye2003}
D.~J. Wales and J.~P.~K. Doye.
\newblock Stationary points and dynamics in high-dimensional systems.
\newblock {\em J. Chem. Phys.}, \textbf{119}:12409--12416, (2003).

\bibitem{CCD}
D.~J. Wales, J.~P.~K. Doye, A.~Dullweber, M.~P. Hodges, F.~Y. Naumkin,
  F.~Calvo, J.~Hern\'andez-Rojas, and T.~F. Middleton.
\newblock The {C}ambridge {C}luster {D}atabase.
\newblock \url{https://www-wales.ch.cam.ac.uk/CCD.html}.

\bibitem{walesdmmw00}
D.~J. Wales, J.~P.~K Doye, M.~A. Miller, P.~N. Mortenson, and T.~R. Walsh.
\newblock Energy landscapes: from clusters to biomolecules.
\newblock {\em Adv. Chem. Phys.}, \textbf{115}:1--111, (2000).

\bibitem{walesmd96}
D.~J. Wales, L.~J. Munro, and J.~P.~K. Doye.
\newblock What can calculations employing empirical potentials teach us about
  bare transition-metal clusters.
\newblock {\em J. Chem. Soc. Dalton Trans.}, \textbf{92}:611--623, (1996).

\bibitem{waless99}
D.~J. Wales and H.~A. Scheraga.
\newblock Global optimization of clusters, crystals and biomolecules.
\newblock {\em Science}, \textbf{285}:1368--1372, (1999).

\bibitem{Whyte}
L.~L. Whyte.
\newblock Unique arrangements of points on a sphere.
\newblock {\em Am. Math. Monthly}, \textbf{59}:606--611, (1952).

\bibitem{WilleVennik}
L.~T. Wille and J.~Vennik.
\newblock Computational complexity of the ground-state determination of atomic
  clusters.
\newblock {\em J. Phys. A: Math. Gen.}, \textbf{18}:L419--422, (1985).

\bibitem{wolters2015self}
J.~R. Wolters, G.~Avvisati, F.~Hagemans, T.~Vissers, D.~J. Kraft, M.~Dijkstra,
  and W.~K. Kegel.
\newblock Self-assembly of “{M}ickey {M}ouse” shaped colloids into
  tube-like structures: {E}xperiments and simulations.
\newblock {\em Soft Matter}, \textbf{11}:1067--1077, (2015).

\bibitem{XJCSinJCP2004}
Y.~Xiang, H.~Jiang, W.~Cai, and X.~Shao.
\newblock An effective method based on lattice construction and the genetic
  algorithm for optimization of large {L}ennard-{J}ones clusters.
\newblock {\em J. Chem. Phys.}, \textbf{108}:3586--3592, (2004).

\bibitem{Xue}
G.~L. Xue.
\newblock Minimum inter-particle distance at global minimizers of
  {L}ennard-{J}ones clusters.
\newblock {\em J. Global Optim.}, \textbf{11}:83--90, (1997).

\bibitem{YWCWinJCP2019}
K.~Yu, X.~Wang, L.~Chen, and L.~Wang.
\newblock Unbiased fuzzy global optimization of {L}ennard-{J}ones clusters for
  $n\leq 1000$.
\newblock {\em J. Chem. Phys.}, \textbf{151}:214105, 1--9, (2019).

\bibitem{Yuhjtman}
S.~A. Yuhjtman.
\newblock A sensible estimate for the stability constant of the
  {L}ennard-{J}ones potential.
\newblock {\em J. Statist. Phys.}, \textbf{160}:1684--1695, (2015).

\bibitem{ZschornakETal}
M.~Zschornak, T.~Leisegang, F.~Meutzner, H.~St\"ocker, T.~Lemser, T.~Tauscher,
  C.~Funke, C.~Cherkouk, and D.~C. Meyer.
\newblock Harmonic principles of elemental crystals --- {F}rom atomic
  interaction to fundamental symmetry.
\newblock {\em Symmetry}, \textbf{10}:228, 1--14, (2018).

\end{thebibliography}

\end{document}